\documentclass[twocolumn]{aastex701}
\usepackage{enumitem}
\newcommand{\Porb}{P_{\mathrm{orb},i}}
\newcommand{\Mhe}{M_{\mathrm{He},i}}
\begin{document}

\title{Shock cooling emission from late-time mass loss in low-mass He star binaries}
\author[0000-0003-2872-5153]{Samantha C. Wu}\thanks{Carnegie Theoretical Astrophysics Center Fellow}
\affiliation{The Observatories of the Carnegie Institution for Science, Pasadena, CA 91101, USA}
\affiliation{Center for Interdisciplinary Exploration \& Research in Astrophysics (CIERA), Physics \& Astronomy, Northwestern University, Evanston, IL 60202, USA}
\email[show]{swu@carnegiescience.edu}  

\author[0000-0001-6806-0673]{Anthony L. Piro}
\affiliation{The Observatories of the Carnegie Institution for Science, Pasadena, CA 91101, USA}
\email{piro@carnegiescience.edu}

\begin{abstract}
     A subset of hydrogen-poor supernovae (SNe) exhibit signatures of interaction with nearby dense circumstellar material (CSM). These SNe may originate from interacting binary systems, in which the SN progenitor experiences intense mass loss when it overflows its Roche lobe close to core collapse. In this work, we explore the appearance of SNe from low-mass stripped star progenitors in binary systems, for a range of initial orbital periods and masses. We model the CSM based on the stripped stars' mass loss history in binary stellar evolution simulations, then numerically explode the progenitors to calculate the SN light curves. Shock cooling emission (SCE) from the CSM dominates the early light curves, followed by SCE from the extended helium envelopes of the stripped stars, which form helium recombination plateaus. The appearance and properties of our model light curves are reflected in a subset of Type Ib/n SNe from the literature. Some of our models tend to evolve rapidly and are quite hot during SCE, so they may naturally explain some fraction of fast blue optical transients (FBOTs).  Since the mass loss history of our binary progenitor models can produce dense CSM out to $\sim 10^{18}\,{\rm cm}$, interaction of this CSM with the SN shock could generate bright late-time radio emission in the years after the optical SN. Searching for late time rising radio emission from FBOTs could be used to test which events are explained by the scenario we explore here.
\end{abstract}

\section{Introduction}

Observations of core-collapse supernovae (SNe) have revealed a population of SNe that show signatures of interaction with dense circumstellar material (CSM). In many cases, the observations indicate that an episode of extreme mass loss towards the end of the SN progenitor's lifetime deposited a large amount of CSM nearby the progenitor star. The SN shock may break out in the CSM, elongating and enhancing the breakout signature \citep{Chevalier2011,Haynie2021} and potentially creating narrow emission lines in early spectra that appear when nearby CSM is flash-ionized by the SN breakout \citep[e.g.,][]{galyam2014,khazov2016}. This shock-heated CSM expands and cools, which manifests as fast-rising, bright early-time optical/UV light curves often accompanied by featureless, blue early spectra \citep[e.g.,][]{Hosseinzadeh2017,pellegrino2022}. 

A subset of these interacting supernovae are classified as Type Ib/c, as their spectra are devoid of hydrogen (H) lines but do show helium (He) lines. Many Type Ib/c exhibit signatures of fast, bright, and blue early emission attributed to shock cooling emission (SCE) from extended helium-rich material \citep[e.g.,][]{De2018,Taddia2018,Ho2020,Jacobson2020,Yao2020,Pellegrino22a}. Some H-poor SNe are characterized by narrow He line emission, called Type Ibn SNe, with typical ejecta masses $M_{\rm ej} = 1\text{--}5\, M_{\odot}$ and $^{56}$Ni masses $M_{\rm Ni} \lesssim 0.1\, M_{\odot}$ \citep[][]{Gangopadhyay2022,maeda2022,ho2021}. Even lower ejecta masses of $M_{\rm ej}\lesssim1\,M_\odot$ are inferred for the ultra-stripped SNe \citep{De2018,Yao2020,yan2023}, which are of particular interest as they may represent explosions from progenitors of compact neutron star binary systems \citep{Dewi2003,Tauris2013,Tauris2015}. In some cases, Type Icn SNe, which interact with He-poor and H-poor material, may share similar origins to ultra-stripped SNe \citep{pellegrino2022b}.

Probable progenitors for interacting H-poor SNe include low-mass stripped stars that have lost their H envelopes through case B mass transfer, which occurs due to Roche lobe overflow after core H burning \citep{smith2017,dessart2022,Ercolino2025}. For stars whose remaining He cores have masses $M_{\rm He} \lesssim 4\, M_{\odot}$, the envelopes expand significantly after core He burning and during core carbon (C) burning, which can lead to Roche lobe overflow in binary systems with separations of less than a few $100\, R_{\odot}$ after case B mass transfer \citep[][]{habets1986,habets1986b,Laplace2020}. The intensity of mass loss during this phase can explain the diversity in the amount of stripping from Type Ib/c progenitors, which retain a more massive He envelope, and ultra-stripped SN progenitors, which may originate from tighter binaries where a large fraction of the He envelope is lost \citep{Yoon2010,Tauris2013,Tauris2015,Zapartas2017,Laplace2020,Wu2022,Ercolino2025}. 

However, this phase of mass transfer occurs too early in the stripped star's lifetime to explain very nearby CSM at $\lesssim 10^{15}$ cm from the progenitor star. Instead, one promising explanation is that another phase of binary interaction begins closer to core collapse, leading to high rates of mass transfer in the last months--years of the star's lifetime \citep{Wu2022}. This could trigger unstable mass transfer and merger \citep{Tsuna2024,dong2024}, but if the mass transfer remains stable, it could lead to dense nearby CSM in the form of He-rich circumbinary material ejected from the binary system. \cite{haynie2025} performed an initial exploration of how SNe from such systems might appear by exploding a few models of stripped stars with CSM originating from late-stage binary mass transfer. These first light curve models showed that subsequent to the SCE from the dense CSM, the SCE from the extended He envelope could be bright enough to appear depending on the explosion energy and $^{56}$Ni mass. Such elevated emission at later epochs qualitatively reproduced the appearance of events such as SN 2019dge \citep{Yao2020}.

In this work, we present a comprehensive exploration of the explosions of low-mass stripped stars that expand significantly before core collapse. We model the evolution of these stripped stars in binaries with varied orbital periods, such that the expansion phase can lead to a range of CSM masses, or no late-time mass loss at all, depending on the initial orbital period. We numerically explode progenitor models of stripped stars with nearby CSM, consistent with the mass loss history in the binary evolution models, and we generate a grid of light curves with different SN explosion energies and masses of $^{56}$Ni. The diverse properties of the resultant light curves range from luminous, rapidly-evolving transients reminiscent of the fastest and brightest Type Ib/c and Type Ibn SNe, to less luminous and gradually evolving light curves dominated by He envelope SCE that are exciting prospects for directly measuring the radii of H-poor SN progenitors.

In Section \ref{sec:methods}, we summarize our methods for constructing our numerical CSM and stripped star progenitors from binary stellar evolution models and computing the light curves from exploding these progenitor models. We review our results for the model light curves in Section \ref{sec:binaryprogenitorLCs} and in Section \ref{sec:observationalcomparisons}, we contextualize our models among observed events via comparisons of the morphology and peak properties of the light curves to known H-poor SNe in the literature. We examine uncertainties, discuss future work, and conclude in Section \ref{sec:discussion}.

\section{Methods} \label{sec:methods}

\subsection{Binary evolution models}
\label{sec:binaryevol}
We use the stellar evolution code MESA \citep[version r15140,][]{mesa2011,mesa2013,mesa2015,mesa2018,mesa2019} to model the binary evolution of stripped stars.
To create each stripped star, we initialize MESA simulations of single stars with initial masses~$M_{\rm ZAMS} = 13.8$--$14.5\, M_{\odot}~$ at $Z = 0.02$, and we remove the entire H envelope once core H burning ends. This leads to stripped stars in the mass range of $\Mhe = 2.5$--$2.75\, M_{\odot}$. Our methods to create and evolve the stripped stars follow those of \cite{Wu2022}. As in previous work, we employ a modified version of the implicit mass transfer scheme of \citet{kolb1990} for Roche lobe overflow that is revised to account for both radiation and gas pressure \citep[e.g.,][]{marchant2021}.

Each stripped star is placed in a circular orbit with a $M_{\rm c} = 1.4\, M_{\odot}$ companion, represented by a point mass in the simulation. In this work, we focus on orbital periods of $\Porb = 30$, $100$, and $300$ d for our binary evolution simulations, with two additional orbital periods of $\Porb = 400$ and $500$ d for the lowest-mass stripped star models that are capable of initiating Roche lobe overflow at larger separations.  We follow the stellar evolution and mass loss of the stripped star from core helium (He) burning up to at least oxygen (O) and neon (Ne) burning, and where possible until silicon burning.

In our models, we assume that CSM is formed via fully non-conservative mass transfer ($f_{\rm mt}=0,\ \beta_{\rm mt}=1$ as in \citealt{mesa2015}), where in MESA we assume that mass and angular momentum are removed from the system in the vicinity of the accretor as a fast wind. For a compact companion star, the mass loss rates in our simulations are many orders of magnitude larger than the Eddington accretion limit of a NS ($\dot{M}_{\rm edd} \sim 4\times 10^{-8}\, M_{\odot}\, \rm{yr}^{-1}$), indicating that most of the mass is likely lost from the system (as in, e.g., \citealt{Tauris2015,Wu2022}); however, for large enough mass loss rates, this process may well take place via the outer Lagrange point L2 \citep{lu2022}. If this is the case, the angular momentum loss of the binary would likely be stronger than what is realized in the models of this work, leading to more rapid orbital decay and higher rates of mass loss. Given that we treat the companion as compact in our simulations, these models most closely describe stripped star+NS systems, but the companion may also represent a low-mass main sequence star. 

As described in detail in \cite{Wu2022}, low-mass He stripped stars with masses of $2.5$--$3\, M_{\odot}$ begin mass transfer after core He exhaustion and during C burning. During the first mass transfer phase from $\sim 10^{3}$--$10^{4}$ yr before core collapse, the mass loss rates reach $\sim 10^{-5}$--$10^{-4}\, M_{\odot}\, \rm{yr}^{-1}$. Depending on the orbital period of the binary and donor star mass, the stripped star may detach from its Roche lobe after the C burning mass transfer phase. Closer to core collapse, the stripped stars tend to expand again during O/Ne burning and initiate intense late-time mass transfer at rates $10^{-3}$--$10^{-1}\, M_{\odot}\, \mathrm{yr}^{-1}$. Table \ref{tab:allmodels} lists the mass-weighted average of the late-time mass transfer rates for each model, $\dot{M}_{\rm csm, avg}$. The amount of radial expansion decreases sharply across this mass range, so stripped stars with larger He core masses undergo weaker mass loss for a given orbital period. In the $\Porb \gtrsim 300$ d simulations, only stripped stars with $\Mhe \lesssim 2.6\, M_{\odot}$ exhibit strong late time mass loss. 

\subsection{Modeling CSM from binary mass transfer}
\label{sec:massloss}

Using our mass loss rates from mass transfer during each binary simulation, $\dot{M}_{\rm csm}$, the density profile of the CSM formed around each binary system is estimated as:
\begin{equation}\label{eq:density}
    \rho_{\rm csm}(r) = \frac{\dot{M}_{\rm csm}}{4\pi r^2 v_{\rm csm}}.
\end{equation}
We expect $v_{\rm csm}$, the velocity of the CSM, to be related to the orbital velocity of the NS companion, $v_{\rm orb, c}$. Smoothed-particle hydrodynamical (SPH) simulations of mass loss from the outer Lagrange point indicate that unbound outflows reach mean asymptotic velocities of a fraction of the binary escape velocity, so that the CSM velocity is approximately \citep{pejcha2016} 
\begin{eqnarray}
    \label{eq:vcsm}
    v_{\rm csm} &=& f_{\infty}\! \sqrt{2} (1+q) v_{\rm orb,c} = f_\infty\sqrt{2}\left[\frac{2\pi G(M_*+M_c)}{P_{\rm orb}}\right]^{1/3}.
\end{eqnarray}
The fraction $f_\infty$ itself also depends on the mass ratio $q=M_{\rm c}/M_{\rm *}<1$, which we estimate using Figure 3 of \cite{pejcha2016}. Using the initial mass of each stripped star, our models are in the range of $q \approx 0.5$--$0.56$ and therefore $f_\infty \approx 0.2$, leading to typical values of $v_{\rm csm}/v_{\rm orb, c} \approx 0.42$--$0.44$.
We estimate the distance that the CSM reaches after leaving the binary system as $r_{\rm csm} = R_{*} + t_{cc}v_{\rm csm}$, where $t_{cc}$ is the time remaining until core collapse when the mass is lost.

Given $\dot{M}_{\rm csm}$, $v_{\rm csm}$, and $r_{\rm csm}$ as functions of time for each binary evolution model, we calculate the CSM density profile $\rho_{\rm csm}(r)$ at the time of SN using Equation (\ref{eq:density}). Note that although we employ a spherically-symmetric formalism, CSM formed from binary interaction may well be asymmetric \citep[][]{pejcha2016,MacLeod2018}. Nevertheless, we proceed with this simplified one-dimensional profile, as our approach to model the light curves also assumes spherical symmetry.

\subsection{Supernova explosion models}
\label{sec:SNEC}
We use the open-source, one-dimensional hydrodynamic radiative transfer code SNEC \citep{SNEC} to model the light curves for each of our stripped star progenitor models.
The progenitors to be exploded consist of the last available stellar profile from each MESA binary simulation, with the CSM profile as formulated in Section \ref{sec:massloss} attached to the exterior.
We excise the inner core of each stellar profile at the Si/O interface, assuming this material will form a neutron star \citep[e.g.,][]{morozova2018}{}. The mass removed is typically $\sim 1.3$--$1.4\, M_{\odot}$, depending on the stellar model. 

As in previous work \citep[e.g.,][]{Haynie2023,haynie2025}, radioactive $^{56} \mathrm{Ni}$ is deposited through the inner 50\% of the stellar ejecta in mass space. The composition profile of the stellar ejecta is smoothed using a ``boxcar'' method with mass width $0.1M_{\odot}$, performed over 4 iterations. Note that the chosen boxcar width is smaller than the nominal value in previous works and in the default version of SNEC, as for these low-mass stripped stars a narrower boxcar mass is warranted to prevent artificially homogenizing the entire ejecta; unfortunately, an approach that is calibrated to observed SNe is not yet available for the stripped-envelope supernova progenitors we consider in this work. In each model, the CSM composition, which consists of He-rich material, is unaltered by this process as we set the mixing to operate only below the boundary between the stellar ejecta and CSM. 

We set the opacity floor to $0.001\, {\rm cm}^2\, {\rm g}^{-1}$, roughly 100 times smaller than the electron scattering value for singly ionized He, which allows for recombination of He without any constraint. We follow \cite{haynie2025} in removing the material above the $\tau=2/3$ photosphere to prevent numerical issues from tracking diffuse material that is rapidly accelerated by the shock. Though the presence of this under-dense material is physically motivated by the MESA binary models of rapid binary mass loss, our choice to remove this mass only minimally impacts both the mass in the system and the appearance of the SCE in the resulting light curves. 

Each SNEC model is exploded using a ``thermal bomb'' mechanism, with final kinetic energies from $E_{\rm SN} = 10^{50}\,{\rm erg}$ to $3\times 10^{51}\,{\rm erg}$, motivated by \cite{Lyman2016}. We also vary the total mass of $^{56} \mathrm{Ni}$ between $M_{\rm Ni} = 0.01$--$0.1\, M_{\odot}$. The simulations are evolved to 100 days beyond the time of explosion, by which time all of the model light curves are dominated by heating from radioactive nickel. 

\subsubsection{Color Temperature}
\label{sec:colortemperature}
Previous work using multigroup radiation hydrodynamics codes such as STELLA \citep{1998ApJ...496..454B,2000ApJ...532.1132B,2006A&A...453..229B,2011ascl.soft08013B} find that the color temperature of photons in stripped-envelope supernovae can exceed the photospheric temperature \citep{jin2023,chiba2025}, motivating an update to the treatment of the observed temperature in SNEC. For example, \cite{chiba2025} find that, upon fitting the spectral energy distribution (SED) of the emitted photons with a blackbody, the inferred color temperature of that blackbody can be 1.5--2 times larger than the photospheric temperature. 

SNEC uses Rosseland mean opacities, but it does estimate the absolute magnitudes in different observed wavelength bands using bolometric corrections, which assume a blackbody at some temperature for the photon SED. Previously, the photospheric temperature, $T_{\rm photo}$, was used to calculate these bolometric corrections. We update SNEC to estimate the color temperature, $T_{\rm color}$, using the methods described below. As a first step toward a more realistic photon energy distribution, we use $T_{\rm color}$ to calculate the flux in observed bands instead.

The photon temperature is likely in thermal equilibrium with the gas temperature for radii within the thermalization radius $r_{\rm th}$, where the thermalization optical depth $\tau_{\rm th} \sim 1$. Here we define
\begin{equation}
\label{eq:tautherm}
    \tau_{\rm th}(r) \approx \int_R^r \rho \sqrt{\kappa_{\rm abs}\kappa_{\rm tot}} \, \mathrm{d}r,
\end{equation}
where $\kappa_{\rm abs}$  is the absorption opacity and $\kappa_{\rm tot} = \kappa_{\rm abs} + \kappa_{\rm scat}$ is the total opacity including the contribution from absorption and scattering ($\kappa_{\rm scat}$). 
The thermalization depth describes the last surface where scattered photons are equally likely to be absorbed. For $\tau_{\rm th} < 1$, photons can escape by scattering without being absorbed, whereas for $\tau_{\rm th} > 1$, photons from deeper layers are mostly absorbed before escaping, and the radiation can approach thermal equilibrium with the matter. 

We estimate the color temperature at each timestep to be the gas temperature at $r_{\rm th}$. At each timestep, we calculate the profile of the thermalization optical depth $\tau_{\rm th}(r)$ and find the location where $\tau_{\rm th}=1$ to identify the color temperature. To estimate the absorption opacity across the grid given the total opacity $\kappa_{\rm tot}$ and the scattering opacity $\kappa_{\rm scat}$, we take $\kappa_{\rm tot}$ to be the Rosseland mean opacity from SNEC, while the scattering opacity is found from $\kappa_{\rm scat} = \sigma_Tn_e/\rho$. The number density of electrons $n_e$ is derived from solving the Saha-Boltzmann equations with the built-in routines of SNEC \citep{SNEC}. Finally, taking the difference yields the absorption opacity, $\kappa_{\rm abs} = \kappa_{\rm tot}-\kappa_{\rm scat}$. 

Our results for the revised color temperatures shown in detail in Appendix Section \ref{sec:appendixcolorTeff}. Overall, we find similar ratios of the color temperature to the photospheric temperature as \cite{chiba2025}, with $T_{\rm color}/T_{\rm photo} \approx 1.2$--$2$ during phases of the light curve when shock cooling emission dominates. We stop tracking the color temperature once the thermalization depth has receded to the innermost grid point in SNEC, which typically occurs once radioactive nickel heating becomes the prevailing source of luminosity. 

\section{Light curves of of binary progenitor models}
\label{sec:binaryprogenitorLCs}
\subsection{Progenitor model properties} 
\label{sec:progenitormodels}
Due to the large radial expansion of these stripped stars during their evolution, the stellar ejecta can extend to $10$s--$100$s of $R_{\odot}$, or $\sim10^{12}$--$10^{13}\, \mathrm{cm}$. The radial extent of the stellar ejecta can depend on the initial orbital period of the binary, as the star is often Roche lobe filling at core collapse. However, some higher-mass stripped stars in wider orbits are not limited by the Roche lobe radius and can expand to the maximum extent based on their isolated evolution. 

The radial extent of the CSM depends strongly on the stripped star initial mass and the orbital period of the binary, as these factors control the time when mass transfer begins. Material located beyond $\gtrsim 10^{15}$ cm tends to be very diffuse and falls below the $\tau=2/3$ photosphere, so it is usually removed from the initial progenitor profile before simulating the explosion. Table \ref{tab:allmodels} lists the outer CSM radius $R_{\rm csm}$ after this cut. Another important radial scale is the shock breakout radius, $R_{\rm sbo}$, which defines the location where $\tau = c/v_s \approx 30 $ (for supernova shock velocities $v_s \approx 10^4\,\mathrm{km}\,\mathrm{s}^{-1}$). This can be written in terms of the diffusion radius $R_d$ \citep{Haynie2021}:
\begin{eqnarray}
    R_{\rm sbo} &=& \frac{R_{\rm csm}}{1+R_{\rm csm}/R_d}\label{eq:rsbo} \\
    R_d &=& \frac{\kappa v_{s} \dot{M}_{\rm{w}}}{4\pi c v_{\rm{w}} }. \label{eq:r_d}
\end{eqnarray}
This assumes a constant mass loss rate for the wind, $\dot{M}_{\rm w}$, lost at a constant wind speed, $v_w$. 

We use the mass-weighted average of the mass loss rate, $\dot{M}_{\rm csm, avg}$, and the wind speed, $v_w = v_{\rm csm}$ (Equation \ref{eq:vcsm}), during the late-time mass loss phase of each He star progenitor to estimate $R_{\rm sbo}$ for each model. 
Our calculation of $R_{\rm sbo}$ is independent of the choice to cut off material above the $\tau=2/3$ photosphere in SNEC. In most models, we find that $R_{\rm sbo}$ is a large fraction of $R_{\rm csm}$, so the shock breaks out near the edge of the CSM and the majority of the CSM contributes to the subsequent shock-cooling emission.

\begin{table*}
\centering
\hspace{-2cm} 
\begin{tabular}{ccccccccc}
\hline
Label & $\Mhe$ $(M_{\odot})$ & $\Porb$ (d) & $M_{\rm ej}$ $(M_{\odot})$  & $M_{\rm csm}$ $(M_{\odot})$ & $R_{\rm ej}$ (cm) &  $R_{\rm csm}$ (cm) &  $R_{\rm sbo}/R_{\rm csm}$ & $\dot{M}_{\rm csm, avg}$ ($M_{\odot}\, \mathrm{yr}^{-1}$) \\
\hline

M2.51P100 & 2.51 & 100 & 0.069 & 0.20 & $6.3\times10^{12}$ & $3.7\times10^{14}$ & 0.72 & $4.2 \times 10^{-2}$  \\
M2.51P300 & 2.51 & 300 & 0.080 & 0.33 & $1.2\times10^{13}$ & $3.0\times10^{14}$ & 0.81 & $4.2 \times 10^{-2}$  \\
M2.51P400 & 2.51 & 400 & 0.30 & 0.85 & $1.5\times10^{13}$ & $2.3\times10^{14}$ & 0.96 & $1.7 \times 10^{-1}$  \\
M2.51P500 & 2.51 & 500 & 0.50 & 0.62 & $1.8\times10^{13}$ & $1.3\times10^{14}$ & 0.97 & $1.6 \times 10^{-1}$  \\

M2.55P30 & 2.55 & 30 & 0.29 & 0.033 & $2.1\times10^{12}$ & $1.5\times10^{14}$ & 0.63 & $1.8 \times 10^{-2}$  \\
M2.55P100 & 2.55 & 100 & 0.24 & 0.24 & $5.4\times10^{12}$ & $1.9\times10^{14}$ & 0.90 & $7.8 \times 10^{-2}$  \\
M2.55P300 & 2.55 & 300 & 0.26 & 0.34 & $1.1\times10^{13}$ & $2.0\times10^{14}$ & 0.92 & $7.7 \times 10^{-2}$  \\

M2.62P30 & 2.62 & 30 & 0.31 & 0.025 & $2.9\times10^{12}$ & $3.5\times10^{13}$ & 0.99 & $1.9 \times 10^{-1}$  \\
M2.62P100 & 2.62 & 100 & 0.58 & 0.095 & $9.8\times10^{12}$ & $3.2\times10^{13}$ & 1.0 & $4.1 \times 10^{-1}$  \\
M2.62P300 & 2.62 & 300 & 1.0 & 0.00 & $8.4\times10^{12}$ & N/A & N/A & N/A  \\

M2.65P30 & 2.65 & 30 & 0.45 & 0.014 & $2.3\times10^{12}$ & $1.0\times10^{13}$ & 1.0 & $1.8 \times 10^{-1}$  \\
M2.65P100 & 2.65 & 100 & 0.50 & 0.12 & $1.0\times10^{13}$ & $6.7\times10^{13}$ & 0.99 & $2.7 \times 10^{-1}$  \\
M2.65P300 & 2.65 & 300 & 1.2 & 0.00 & $8.7\times10^{12}$ & N/A & N/A & N/A  \\

M2.72P30 & 2.72 & 30 & 0.71 & 0.0078 & $2.0\times10^{12}$ & $1.4\times10^{13}$ & 0.99 & $1.3 \times 10^{-1}$  \\
M2.72P100 & 2.72 & 100 & 0.98 & 0.022 & $6.3\times10^{12}$ & $1.3\times10^{13}$ & 1.0 & $2.7 \times 10^{-1}$  \\
M2.72P300 & 2.72 & 300 & 1.3 & 0.00058 & $9.7\times10^{12}$ & $1.4\times10^{13}$ & 0.89 & $5.1\times10^{-3}$  \\

M2.75P30 & 2.75 & 30 & 0.90 & 0.020 & $2.3\times10^{12}$ & $1.7\times10^{13}$ & 0.99 & $2.1 \times 10^{-1}$  \\
M2.75P100 & 2.75 & 100 & 1.2 & 0.0032 & $4.8\times10^{12}$ & $1.3\times10^{13}$ & 0.97 & $3.0 \times 10^{-2}$  \\
M2.75P300 & 2.75 & 300 & 1.3 & 0.00 & $6.8\times10^{12}$ & N/A & N/A & N/A  \\

\hline \\[-1mm]
\end{tabular}
\caption{Properties of selected models presented throughout this work. Listed values include the initial mass of each He star progenitor, $\Mhe$; the initial orbital period of the binary progenitor system, $\Porb$; the ejecta mass of the He star in the SNEC model, $M_{\rm ej}$; the total mass of CSM in the SNEC model, $M_{\rm csm}$; the initial radius of the He star envelope, $R_{\rm ej}$; and the total radial extent of the CSM in the SNEC model, $R_{\rm csm}$. In the SNEC model, $R_{\rm csm}$ is set by the location of the $\tau=2/3$ photosphere, as material above is removed from the SNEC grid; this choice minimally impacts $M_{\rm csm}$. We also estimate the shock breakout radius $R_{\rm sbo}$ assuming a constant $\dot{M}$ wind (Equation \ref{eq:rsbo}), using a mass-weighted average of the mass loss rate, $\dot{M}_{\rm csm, avg}$, and the wind speed $v_w = v_{\rm csm}$ (Equation \ref{eq:vcsm}) during the late-time mass loss phase for each He star progenitor in MESA. The values of $R_{\rm sbo}$/$R_{\rm csm}$ and $\dot{M}_{\rm csm, avg}$ are listed in the final two columns of the table.}
\label{tab:allmodels}
\end{table*}

Shock cooling emission (SCE) is an important component of our model light curves. The approximate timescale of the SCE phase can be understood from the diffusion time, $t_d$: 
\begin{equation}
    t_d \propto \left(\frac{\kappa M_{\rm ext}^{3/2}}{E_{\rm ext}^{1/2}c}\right)^{1/2}.
    \label{eq:td}
\end{equation}
The luminosity of the SCE phase at the time $t=t_d$ goes as \citep{piro2021}
\begin{equation}
    L_{\rm SCE} (t_d) \propto \frac{c R_{\rm ext}}{\kappa}\frac{E_{\rm ext}}{M_{\rm ext}}.
    \label{eq:lumSCE}
\end{equation}
In the above, $R_{\rm ext}$ and $M_{\rm ext}$ refer to the radius and mass of the extended material that is producing the SCE, and $E_{\rm ext}$ describes how much energy the shock transfers into the extended material. $E_{\rm ext}$ is related to $M_{\rm ext}$, the ejecta mass $M_{\rm ej}$, and the total kinetic energy of the supernova $E_{\rm SN}$ by \citep{Nakar2014}
\begin{equation}
    E_{\rm ext} = 4\times10^{49}\, {\rm erg} \left(\frac{E_{\rm SN}}{10^{51}\,{\rm erg}}\right)
     \left(\frac{M_{\rm ext}}{0.01\,M_\odot}\right)^{0.7}
    \left(\frac{M_{\rm ej}}{1\,M_\odot}\right)^{-0.7},
\end{equation}
assuming that $M_{\rm ext} < M_{\rm ej}$.

As explored in detail in \cite{haynie2025}, both the CSM and the extended He envelope of these binary stripped star progenitors can contribute to observable phases of SCE. Table \ref{tab:allmodels} lists the mass of the CSM, $M_{\rm csm}$, and mass of the extended He envelope, which we take to be equal to the ejecta mass $M_{\rm ej}$, in each progenitor model. For the CSM phase of SCE, Equations (\ref{eq:td}) and (\ref{eq:lumSCE}) can be evaluated using $R_{\rm ext} \approx R_{\rm sbo}\approx R_{\rm csm}$, $M_{\rm ext}\approx M_{\rm csm}$, and $M_{\rm ej}$. For the SCE phase from the extended He envelope, the approximate SCE timescale and an upper limit on $L_{\rm SCE}$ can be estimated using $R_{\rm ext} \approx R_{\rm ej}$ and $M_{\rm ext}\approx M_{\rm ej}$, while also assuming $E_{\rm ext} \approx E_{\rm SN}$. 

The shock cooling emission from CSM is always brighter than the SCE from the extended He envelope due to the larger radius of the CSM. However, \cite{haynie2025} demonstrated that the SCE from the extended He envelope can contribute to the light curve on longer timescales than the CSM-dominant phase of SCE, as the He envelope mass can be much larger than the CSM mass. Our stripped star models, which expand significantly during late nuclear burning stages, remain very extended until core collapse. As a result, the He envelope radii in our models are large enough that the luminosity from the He envelope phase of SCE could enhance the light curve significantly above the radioactive decay of $^{56}$Ni on these timescales, depending on the explosion energy.

Among our progenitor models, we see that models originating from smaller initial orbital periods $\Porb$ have smaller He envelope radii $R_{\rm ej}$. When $\Porb$ is small enough that the stripped star is filling its Roche lobe at core collapse, the He envelope radii are limited to the size of the Roche lobe. At larger $\Porb$ where the expansion is not limited by the Roche lobe, lower-mass models can achieve larger radii than higher-mass models, and the He envelope SCE of higher-mass models is thus expected to be less bright. 

The He envelope masses, or equivalently ejecta masses, vary with orbital period and mass as well. Models that have shorter orbital periods will have lost more mass during the C burning mass transfer phase occurring $\approx 10^4$ yr before core collapse, leading to smaller ejecta masses for a given $\Mhe$. Similarly, as lower-mass stripped stars rapidly expand to large maximum radii, they tend to overflow their Roche lobes earlier before core collapse than higher-mass stripped stars at a given orbital period, and therefore have a longer time interval for mass loss. This leads to smaller final ejecta masses for lower-mass stripped stars at fixed $\Porb$.

The M2.51P400 and M2.51P500 models place the lowest-mass stripped star progenitor of $\Mhe=2.51\, M_{\odot}$ in orbits with large values of $\Porb = 400$ d and $\Porb = 500$ d. This allows the progenitor to avoid transferring any mass during the C burning mass transfer phase, but instead lose $\sim 0.5$--$1\, M_{\odot}$ of CSM during a late-time mass transfer phase. The ejecta masses are comparable to the CSM masses due to this mass loss history, leading to similar diffusion timescales for both SCE phases. As a result, we expect the combined SCE phases to form an extended plateau in the light curve \citep{haynie2025}. Similar arguments apply to the M2.55P300 and M2.55P100 models, which also have nearly equivalent values of $M_{\rm csm}$ and $M_{\rm ej}$.

\begin{figure}
    \flushleft
    \includegraphics[width=0.99\columnwidth]{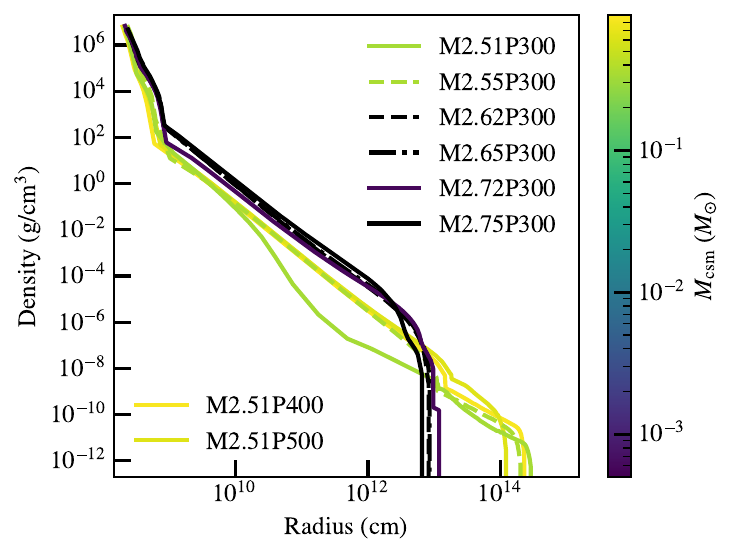}
    \includegraphics[width=0.99\columnwidth]{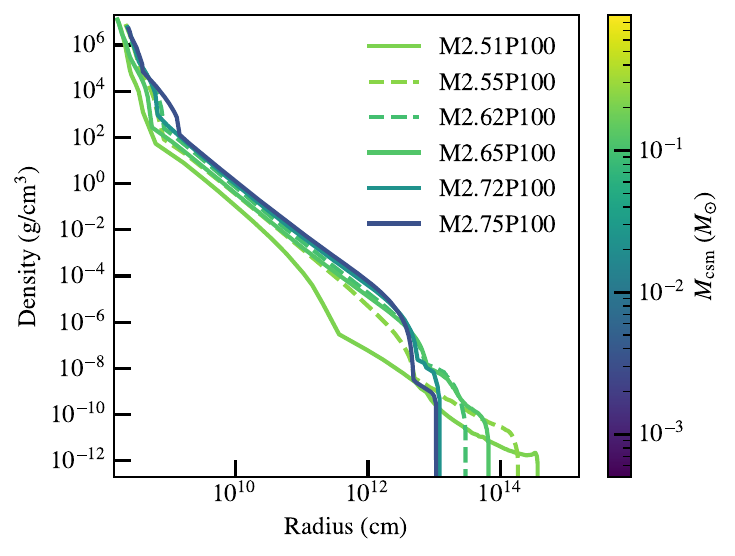}
    \includegraphics[width=\columnwidth]{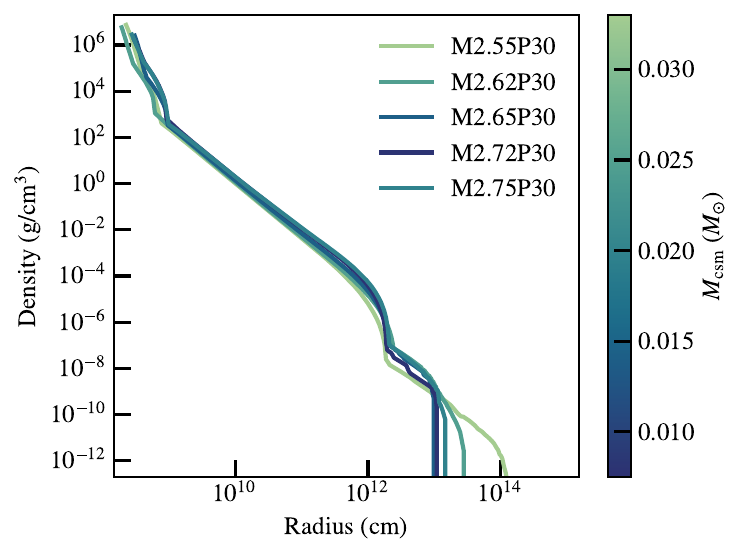}
    \caption{Model density profiles. The top panel shows models with $\Porb = 300$ d, along with $\Porb=400$ d and $\Porb=500$ d for the $\Mhe=2.51\, M_{\odot}$ model. The middle panel shows the models with $\Porb$ = 100 d, and the bottom panel shows models with $\Porb$ = 30 d. The colors of each profile indicate the mass of the CSM, as indicated in the color bar. Note the different color bar scaling for the $\Porb$ = 30 d models in the bottom panel.}
    \label{fig:allmodeldensityprofs}
\end{figure}

Figure \ref{fig:allmodeldensityprofs} shows the density profiles of each of the progenitor models for which we calculate light curves using SNEC. The density profiles are colored by the mass of the CSM in the model; note that the top two panels have the same color bar scaling, but the third panel for the $\Porb=30$ d models is limited to a smaller dynamic range. Models colored with black lines in the $\Porb=300$ d panel represent extended stripped stars without significant CSM ($M_{\rm csm}\lesssim 10^{-5}\, M_{\odot}$). 

The $\Porb=30$ d models all have similar density profiles characterized by a denser He envelope out to $\sim 10^{12}$ cm surrounded by material of mass $M_{\rm csm} \sim 10^{-2}\, M_{\odot}$, distributed as a $\rho \propto r^{-2}$ wind. Note that we do not include a progenitor model for $\Mhe = 2.5\, M_{\odot}$, $\Porb=30$ d due to numerical issues. The radial extent of each profile varies, influenced both by the time before core collapse when mass transfer begins and by the CSM velocity. 

For the models at larger $\Porb$, the density profiles generally exhibit these same components, but with greater variation in the CSM masses and radial extents. The lowest-mass $\Mhe = 2.5\, M_{\odot}$ model has very low ejecta mass $M_{\rm ej} \lesssim 0.1\, M_{\odot}$ for $\Porb=100$ d and  $\Porb=300$ d, so the He envelope is much less dense than for the other progenitor models. Models without significant CSM have density profiles that drop off sharply at the edge of each stripped star progenitor's envelope, but this is already quite radially extended, reaching $\sim 10^{13}$ cm.

\begin{figure*}
    \centering
    \includegraphics[width=0.32\textwidth]{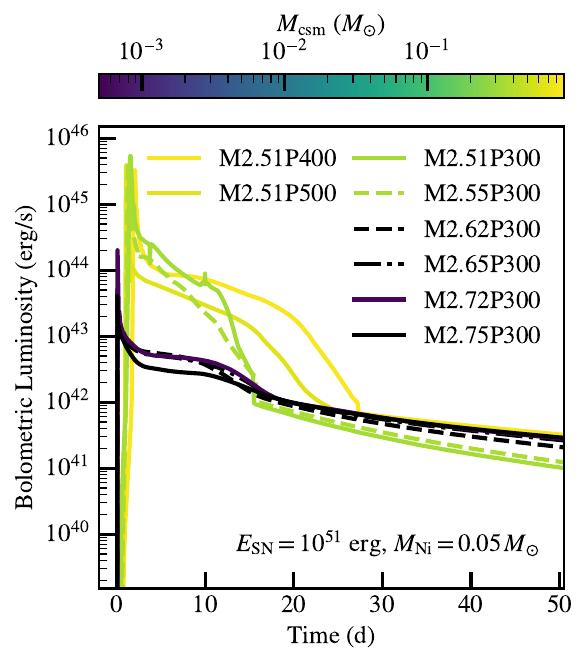}
    \includegraphics[width=0.32\textwidth]{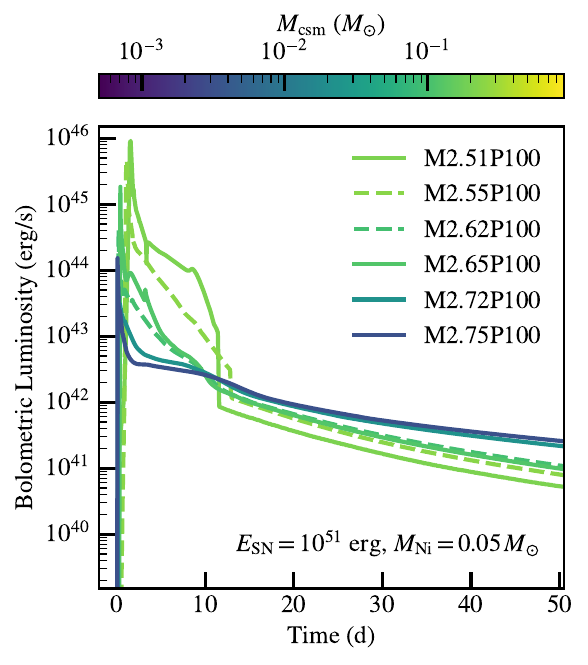}
    \includegraphics[width=0.32\textwidth]{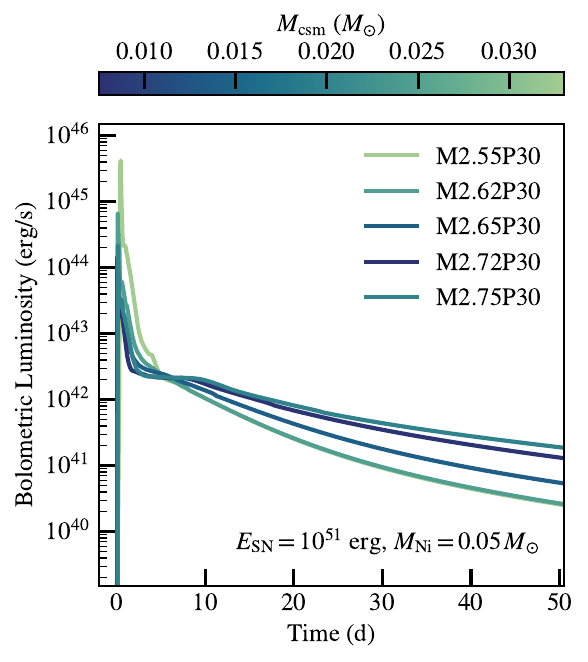}
    \includegraphics[width=0.32\textwidth]{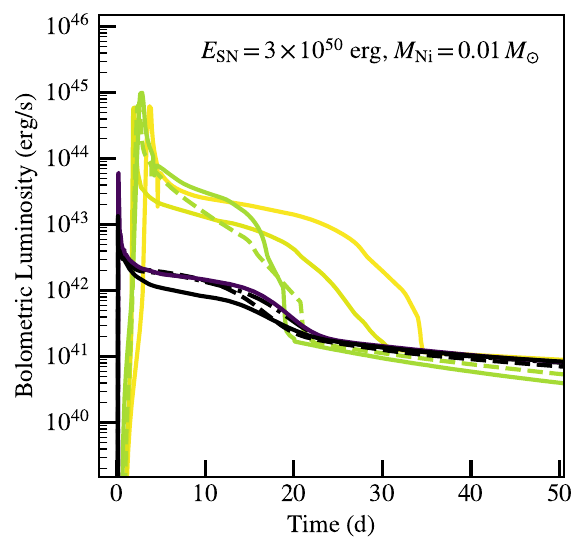}
    \includegraphics[width=0.32\textwidth]{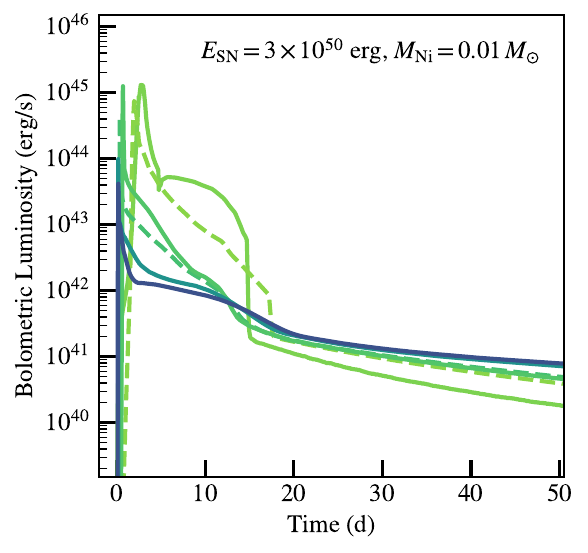}
    \includegraphics[width=0.32\textwidth]{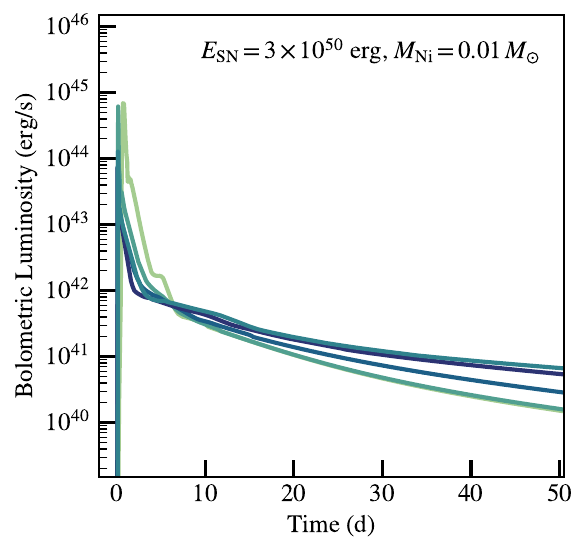}
    \caption{Model light curves for explosion parameters of $E_{\rm SN} = 10^{51}\, \mathrm{erg}$, $M_{\rm Ni}=0.05\, M_{\odot}$ in the top row, and explosion parameters of $E_{\rm SN} = 3\times10^{50}\, \mathrm{erg}$, $M_{\rm Ni}=0.01\, M_{\odot}$ in the bottom row. As in Figure \ref{fig:allmodeldensityprofs}, the line colors indicate the CSM mass; note the different color bar scaling for the $\Porb$ = 30 d models in the right panels.}
    \label{fig:fiducialLCs}
\end{figure*}

\subsection{Fiducial model light curves}
\label{sec:fiducialLCs}
The top row of Figure \ref{fig:fiducialLCs} shows the suite of model light curves for the parameters $E_{\rm SN}=10^{51}$ erg, $M_{\rm Ni}=0.05\, M_{\odot}$. Each of the model light curves begins with shock breakout, which is brighter and lasts longer for models with more massive CSM \citep{Haynie2021,haynie2025}. For some models, the shock breakout width is likely too narrow to be well-resolved by the time resolution of the SNEC simulations, so the actual shock breakout could be even brighter than shown in Figure \ref{fig:fiducialLCs}. Nevertheless, the light curves are mostly shaped by SCE, which we do robustly resolve.

Some models exhibit extremely bright phases of SCE that can reach peak bolometric luminosities of $\gtrsim 10^{44}\, \mathrm{erg}\, \mathrm{s}^{-1}$ and last for $\gtrsim 10$ d.  These light curves come from models with $M_{\rm csm}\gtrsim 0.1\, M_{\odot}$ (yellow and green curves in the left and middle panels), which correspond to the lowest initial He masses of $\Mhe \leq 2.65\, M_{\odot}$ at $\Porb = 100$ d and  $\Mhe \leq 2.55\, M_{\odot}$ at $\Porb \geq 300$ d. 
As expected, the light curve gently declines throughout a plateau lasting $\sim 20$--$30$ d for each of the M2.51P400 and M2.51P500 models, a feature which arises from the similar photon diffusion timescales for both the CSM and He envelope SCE phases. After the SCE ends in these models, the light curve drops abruptly to the radioactive decay tail. In the other models with massive CSM, the light curves also demonstrate elevated plateaus from the combined SCE that decline more steeply.

For the highest mass $\Porb=100$ d models with lower $M_{\rm csm}\sim 10^{-2}\, M_{\odot}$, the initial SCE from the CSM lasts only a few days before transitioning to the lower-luminosity SCE from the He envelope, which appears as a gentle decline from $L_{\rm bol}\sim 3 \times 10^{42}\, \mathrm{erg}\, \mathrm{s}^{-1}$ over $\sim 10$ d. The $\Porb=30$ d models also have $M_{\rm csm}\sim 10^{-2}\, M_{\odot}$ and therefore similar early light curves, but the He envelope SCE is less conspicuous in the $\Porb=30$ d  model light curves. As these stars were placed in tighter orbits, they have smaller envelope radii $R_{\rm ej}$, resulting in less luminous SCE. 

In the left panel, the dark purple and black light curves demonstrate how explosions of low-mass stripped stars without significant CSM still exhibit visible SCE from the extended He envelope. As these stars are not Roche-lobe filling at core collapse, the He envelopes are at their maximum extents in these models. With envelope masses of $M_{\rm ej} \approx 1\, M_{\odot}$, the SCE phase in each light curve appears as a plateau over $10$ d. The SCE luminosities vary slightly depending on the radii of the stripped stars, which range between $6$--$9 \times 10^{12}$ cm. 

The bottom row of Figure \ref{fig:fiducialLCs} shows light curves for lower explosion energy and $^{56}$Ni mass, $E_{\rm SN}=3\times10^{50}$ erg, $M_{\rm Ni}=0.01\, M_{\odot}$. These choices may be more appropriate for low-mass He star progenitors, based on correlations for core-collapse supernovae \citep{Lyman2016,Burrows2024}. With the decreased explosion energy, the longer timescales of SCE lead to broader light curve plateaus with lower luminosities. Once SCE is no longer dominant, the radioactive decay tail is also slightly dimmer with the lower $^{56}$Ni mass.
Across our grid of models, the peak of the light curve is dominated by SCE, so changes in the explosion energy significantly impact the peak luminosity and timescale of the light curves (see Appendix \ref{appendix:energy}, Figure \ref{fig:EvariedLCs}). In contrast, changes in the $^{56}$Ni mass do not greatly affect the maximum luminosity achieved by SCE, but the shape of the SCE bump does appear more prominent in the light curve morphology for lower $M_{\rm Ni}$ at fixed $E_{\rm SN}$ (see Appendix \ref{appendix:Ni}, Figure \ref{fig:NivariedLCs}).

\subsection{Helium shock-cooling plateau luminosity}
\label{sec:HeSCEplateau}

\begin{figure}
    \centering
    \includegraphics[width=\columnwidth]{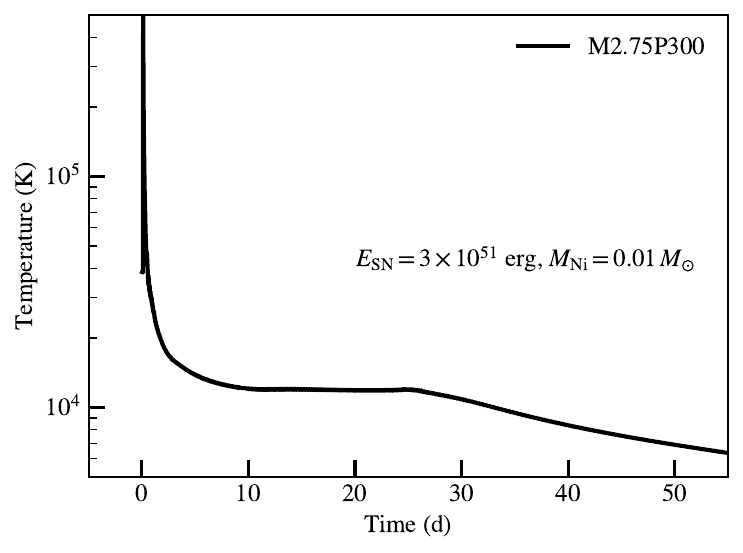}
    \caption{Color temperature evolution in the M2.75P300 model at $E_{\rm SN}=3\times10^{50}$ erg, $M_{\rm Ni}=0.01\, M_{\odot}$. Helium recombination during the shock cooling emission of the extended He envelope causes an opacity jump that sets the thermalization depth at a constant color temperature. Once the helium has mostly recombined, the temperature declines. }
    \label{fig:temperature}
\end{figure}

Figure \ref{fig:temperature} shows an example of the color temperature evolution for M2.75P300 at $E_{\rm SN}=3\times10^{50}$ erg, $M_{\rm Ni}=0.01\, M_{\odot}$. This model has no CSM, but an extended He envelope with $M_{\rm ej} =1.3\, M_{\odot}$ that produces a plateau in the bolometric luminosity lasting $\approx 20$ d. During the He envelope SCE phase, the color temperature remains quite constant around $10^4$ K. We find that the location of the thermalization depth (Equation \ref{eq:tautherm}) throughout this phase is mediated by the drop in opacity due to He recombination at $\sim 10^4$ K. 
The color temperature then begins to decline when the helium has mostly recombined. 

This behavior is analogous to the H recombination plateau in Type IIP supernovae \citep[e.g.,][]{kasen2009}, and has been discussed in the context of the explosions of Wolf-Rayet stars \citep[e.g.,][]{ensman1988,dessart2011,kleiser2014}; it is also representative of the physics underlying all of our models during He SCE, as long as the SCE phase from CSM does not overlap with the He envelope SCE. 
As a result, we can apply similar arguments to the plateaus in our model light curves as have been outlined for Type IIP SNe.

In Type IIP SNe, H recombination creates a sharp ionization front, such that material below the front is opaque and material above is optically thin, and the photosphere of the SN is set at the location of the ionization front \citep{kasen2009}. The plateau lasts until the entire H envelope is recombined, then the light curve sharply drops to the radioactive tail. Analytic scalings for the luminosity and duration of the recombination plateau depend on the energy of the SN $E_{\rm SN}$, the ejecta mass $M_{\rm ej}$, the initial radius of the ejecta $R_0$, the opacity of the ejecta $\kappa$, and the ionization temperature $T_\mathrm{I}$ \citep{popov1993,kasen2009}:
\begin{eqnarray}
    \label{eq:typeIIPscalings_t}
    t_{\rm pl} &\propto& E_{\rm SN}^{-1/4} M_{\rm ej}^{1/2} R_0^{1/6} \kappa^{1/6} T_{\mathrm{I}}^{-2/3} \\
    \label{eq:typeIIPscalings_L}
    L_{\rm pl} &\propto& E_{\rm SN}^{5/6} M_{\rm ej}^{-1/2} R_0^{2/3} \kappa^{-1/3} T_{\mathrm{I}}^{4/3}         
\end{eqnarray}
where $t_{\rm pl}\propto E_{\rm SN}^{-1/4}$ comes from the revised scaling in \cite{kasen2009} based on model light curves of SNe IIP. 

\begin{figure}
    \raggedleft
    \includegraphics[width=0.972\columnwidth]{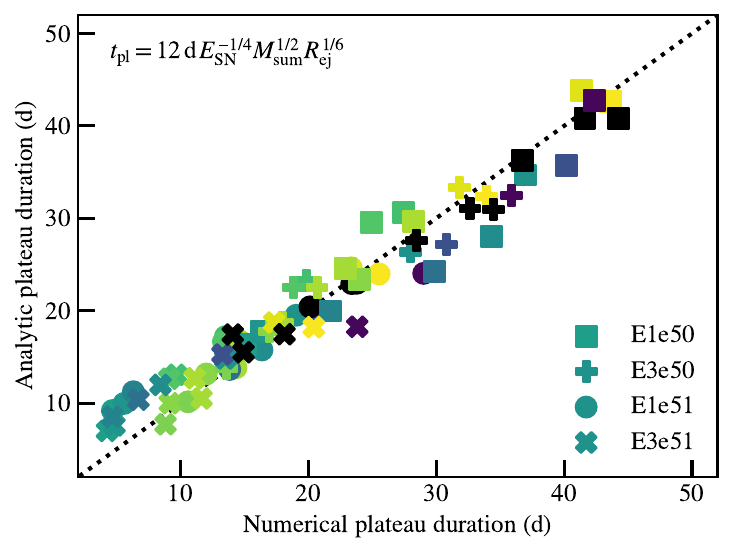}
    \includegraphics[width=\columnwidth]{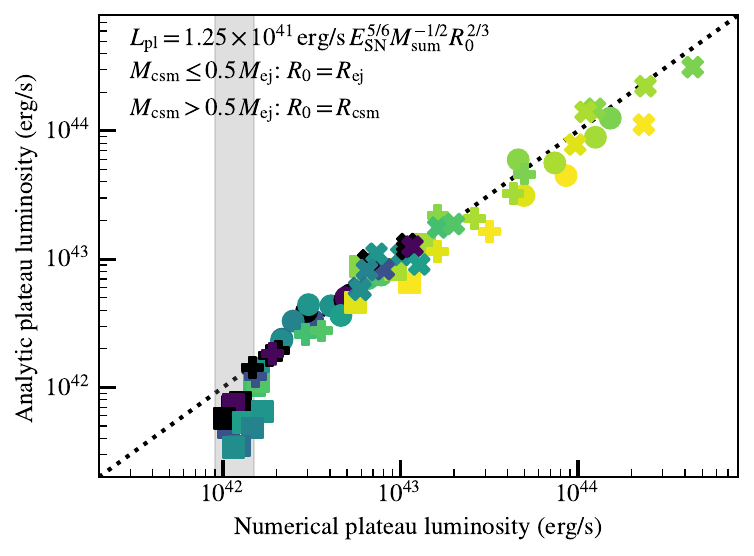}
    \includegraphics[width=\columnwidth]{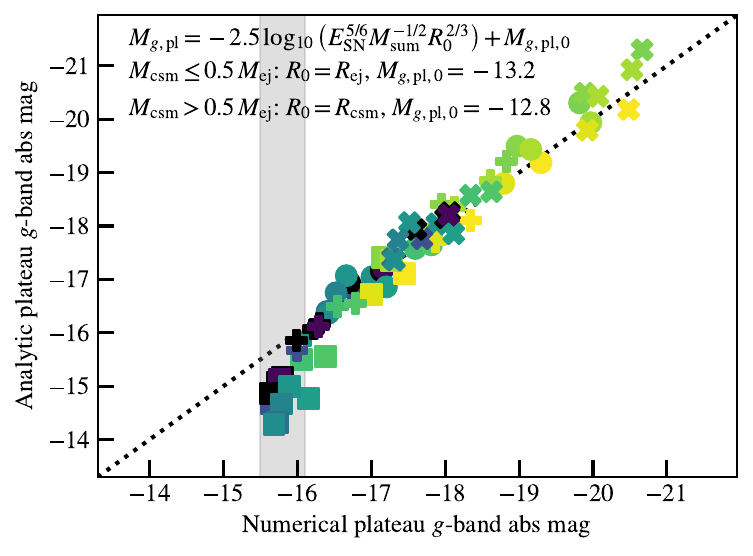}
    \caption{Accuracy of the analytic scalings in Equations (\ref{eq:typeIIPscalings_t}) and (\ref{eq:typeIIPscalings_L}) compared with the SCE plateau in the numerical models. From top to bottom, we show the plateau duration, the bolometric plateau luminosity, and an estimate of the $g$-band absolute magnitude. The analytic predictions are evaluated using the expressions written in each panel.
    Colors represent $M_{\rm csm}$, as in Figure \ref{fig:allmodelprops_obscomparison}, and shapes denote $E_{\rm SN}$. All models use $M_{\rm Ni}=0.05\, M_{\odot}$. The range in contribution from $^{56}$Ni decay to the numerical luminosity is shown as a grey shaded region.  }
    \label{fig:plateau_props_comparison}
\end{figure}

For the H recombination plateau, the sudden change in opacity occurs at ionization temperature $T_I\approx 6000$ K. For He-rich material, the opacity cliff corresponds to a higher temperature of $T_I\approx 1$--$1.5 \times 10^4$ K, when singly ionized He II recombines to form He I \citep{farag2024}. Recombination of fully ionized He into singly ionized He II does create a bump in the opacity at higher temperatures $\sim 5\times 10^4$ K, but the significant opacity drop and recombination plateau occurs at lower temperatures when He becomes fully neutral. In our calculations of the He plateau analytic relations, we assume $T_I \approx 10^4$.

We evaluate these scalings against the properties of the He SCE plateaus in our models in Figure \ref{fig:plateau_props_comparison}. To better reflect the amount of mass in our models, we use $M_{\rm sum}=M_{\rm ej} + M_{\rm csm}$ in place of $M_{\rm ej}$ in Equations (\ref{eq:typeIIPscalings_t}) and (\ref{eq:typeIIPscalings_L}). 
All models shown here are exploded with $M_{\rm Ni}=0.05\, M_{\odot}$. However, the explosion energy is varied from $E_{\rm SN}=10^{50}\, \mathrm{erg}$ to $E_{\rm SN}=3\times10^{51}\, \mathrm{erg}$, with different energies represented by different scatter point shapes in the figure. The numerical plateau luminosity is reported at times before the downturn in the bolometric luminosity plateau, which ranges from $\sim 5$--$10$ d depending on $E_{\rm SN}$. 

For the plateau duration, we compare to the scaling 
\begin{equation}
    \label{eq:tpl_Msun_Rej}
    t_{\rm pl} = 12\, \mathrm{d}\, E_{\rm SN}^{-1/4} M_{\rm sum}^{1/2} R_{\rm ej}^{1/6}.
\end{equation}
Here $E_{\rm SN}$ is in units of $10^{51}$ erg, $M_{\rm sum}$ is in units of $M_{\odot}$, and $R_{\rm ej}$ is in units of $R_{\odot}$.
The top panel shows that with some scatter, the analytic scaling describes the plateau durations $t_{\rm pl}$ of the models fairly well. 

For the plateau luminosity, we compare the models to two different relations. For models with $M_{\rm csm} \lesssim 0.5 M_{\rm ej}$, the plateau luminosity is best described by:
\begin{equation}
    \label{eq:Lpl_Rej}
    L_{\rm pl} =1.25\times 10^{41}\, \mathrm{erg}\,\mathrm{s}^{-1}\,E_{\rm SN}^{5/6} M_{\rm sum}^{-1/2} R_{\rm ej}^{2/3}. 
\end{equation}
For models with $M_{\rm csm} \gtrsim 0.5 M_{\rm ej}$, we compare to the same equation, but with $R_{\rm csm}$ instead of $R_{\rm ej}$:
\begin{equation}
    \label{eq:Lpl_Rcsm}
    L_{\rm pl} =1.25\times 10^{41}\, \mathrm{erg}\,\mathrm{s}^{-1}\,E_{\rm SN}^{5/6} M_{\rm sum}^{-1/2} R_{\rm csm}^{2/3}. 
\end{equation}
The SCE from the CSM blends with the SCE plateau from the He envelope for these models, so the measured plateau luminosity from the numerical light curve is much higher than the predicted analytic plateau luminosity using $R_0=R_{\rm ej}$. Nevertheless, the light curves for the group of models with $M_{\rm csm} \gtrsim 0.5 M_{\rm ej}$ are governed by similar physics as the rest of the models and can be explained by the same analytic scalings, but using the larger radius of $R_{\rm csm}$ instead. 

In the middle panel of Figure \ref{fig:plateau_props_comparison},  we compare the numerical plateau luminosity to the analytic predictions from Equations (\ref{eq:Lpl_Rej})--(\ref{eq:Lpl_Rcsm}). 
The dotted line plots the 1:1 relation between the numerical and analytic values. 
We see the plateau luminosity is well-described by the analytic scalings above $L_{\rm pl}\sim 2\times 10^{42}\, \mathrm{erg}\,\mathrm{s}^{-1}$. However, the numerical plateau luminosities tend to be brighter than the analytic plateau luminosities below that value. 

The models that diverge from the 1:1 relation at low plateau luminosities come from $E_{\rm SN} \lesssim 3\times 10^{50}$ erg explosions. At these lower energies, the contribution to the bolometric luminosity from the radioactive decay of $^{56}$Ni becomes comparable to that of the SCE. This sets a lower limit to the numerical plateau luminosity of at least $\sim 10^{42}\,  \mathrm{erg}\,\mathrm{s}^{-1}$, and the SCE luminosity adds an additional component of comparable magnitude. This causes the numerical plateau luminosity to be systematically larger than the analytic plateau luminosity below $L_{\rm pl}\sim 2\times 10^{42}\, \mathrm{erg}\,\mathrm{s}^{-1}$. We show the typical range of the contribution from radioactive $^{56}$Ni decay as a gray shaded region in the middle panel of Figure \ref{fig:plateau_props_comparison}.



The scalings in Equations (\ref{eq:typeIIPscalings_t}) and (\ref{eq:typeIIPscalings_L}) reference the bolometric light curves, but we can also use our numerical models to infer the relationship between the absolute magnitude on the plateau and the explosion properties. 
For models with $M_{\rm csm} \lesssim 0.5 M_{\rm ej}$, we estimate the absolute magnitude in $g$-band during the He SCE to be approximately 
\begin{equation}
    \label{eq:Mgpl_Rej}
    M_{g,\rm pl}=-2.5\, \log_{10}\left( E_{\rm SN}^{5/6} M_{\rm sum}^{-1/2} R_{\rm ej}^{2/3} \right) - 13.2, 
\end{equation}
Models with $M_{\rm csm} \gtrsim 0.5 M_{\rm ej}$ can be described using
\begin{equation}
    \label{eq:Mgpl_Rcsm}
    M_{g,\rm pl}=-2.5\, \log_{10}\left( E_{\rm SN}^{5/6} M_{\rm sum}^{-1/2} R_{\rm csm}^{2/3} \right) - 12.8, 
\end{equation}
again using the larger radius of $R_{\rm csm}$ instead of $R_{\rm ej}$ to capture the strong influence from CSM SCE. In addition, these models require a slightly dimmer normalization because the higher temperature during the CSM SCE lowers the $g$-band magnitude in the models with $M_{\rm csm} \gtrsim 0.5 M_{\rm ej}$. 

The bottom panel of Figure \ref{fig:plateau_props_comparison} shows the accuracy of these analytic estimates for our numerical models, with the 1:1 relation between the numerical $g$-band plateau brightness for these models and the analytic predictions of Equations (\ref{eq:Mgpl_Rej}) and (\ref{eq:Mgpl_Rcsm}) shown as a dotted black line.
We find that the Ni luminosity enhances the $g$-band plateau brightness between $M_{g,\rm pl}\approx-15.5$ and $M_{g,\rm pl}\approx-16$. 

Overall, the He SCE plateau properties may be used to approximately infer the He envelope initial radius and ejecta mass. Given an independent estimate of the explosion energy,  the relations listed in the top left of each panel in Figure \ref{fig:plateau_props_comparison} would allow for calculation of the total amount of mass in the expanding ejecta $M_{\rm sum} = M_{\rm ej}+M_{\rm csm}$, as well as the initial envelope radius of the stripped star progenitor, $R_{\rm ej}$. The velocity of the SN can often provide an estimate of the ratio $E_{\rm SN}/M_{\rm ej}$, so combined with a measurement of the plateau luminosity and duration for these light curves, these analytic relations can be solved for the unknowns $M_{\rm sum}$ and $R_{\rm ej}$. The He SCE phase during these SN explosions therefore provides an excellent opportunity to estimate the properties of the progenitor system. 

\section{Comparison to observed events}
\label{sec:observationalcomparisons}
\subsection{Bolometric light curve morphology}
\label{sec:bolometriccomparisons}

\begin{figure}
    \centering
    \includegraphics[width=\columnwidth]{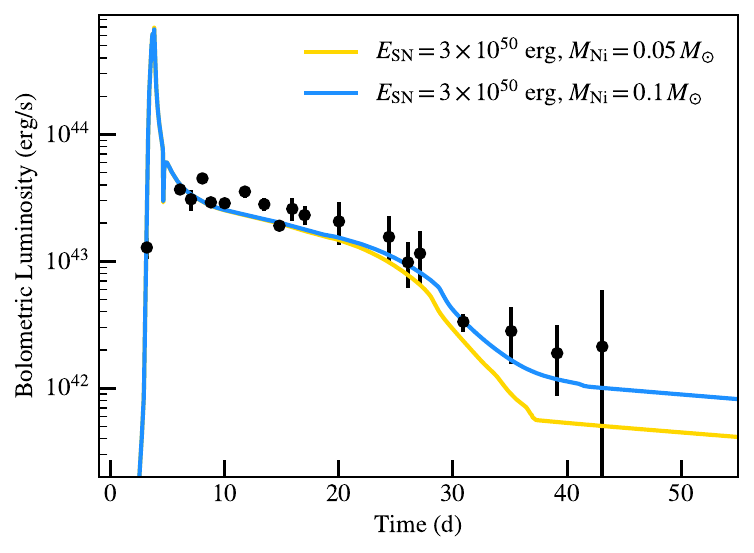}
    \caption{Comparison of the M2.51P400 model with the inferred bolometric luminosity for SN 2019kbj, a Type Ibn SN \citep{Ben-Ami2023}. The model is shown for two choices of the explosion energy $E_{\rm SN}$ and $^{56}$Ni mass $M_{\rm Ni}$ as listed in the legend, and the observed data is shown as black scatter points.}
    \label{fig:sn2019kbj}
\end{figure}

Comparison of our model light curves to observations of rapidly-evolving H-poor SNe reveal that our models resemble various light curve shapes seen in the literature. For instance, Figure \ref{fig:sn2019kbj} shows an example of how the elevated bolometric luminosity during SCE in our M2.51P400 model is able to explain a similar inferred plateau in SN 2019kbj, which is a Type Ibn SN \citep{Ben-Ami2023}. \cite{haynie2025} also compared to this event but were unable to match the full extent of the observed plateau, which suggested a larger He envelope mass. Indeed, our M2.51P400 model has a more massive He envelope of $\approx 0.3\, M_{\odot}$ due to the large initial orbital period of the model, which allows the stripped star to avoid an earlier phase of mass transfer (see Section \ref{sec:progenitormodels}). Using $M_{\rm Ni}=0.1\, M_{\odot}$ to match the late-time tail, the M2.51P400 model with $E_{\rm SN} = 3\times10^{50}$ erg replicates the observed data well. 

\begin{figure}
    \centering
    \includegraphics[width=\columnwidth]{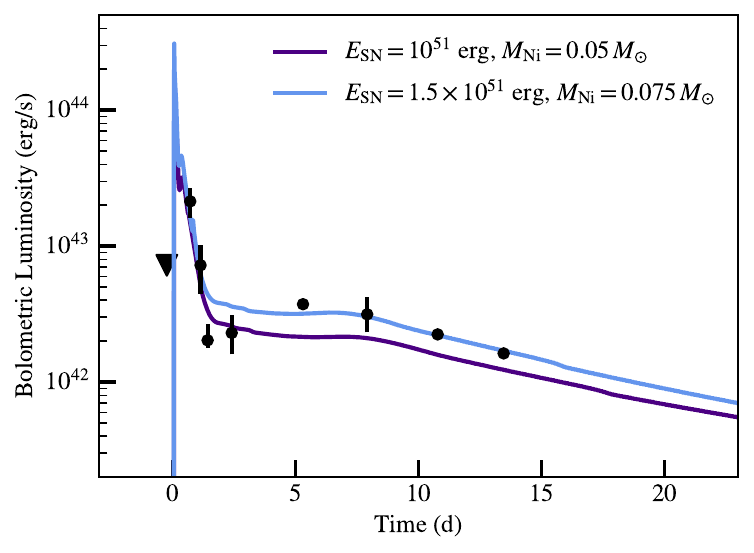}
    \caption{Comparison of the M2.72P30 model with the inferred bolometric luminosity for iPTF14gqr, an ultra-stripped SN \citep{De2018}. The model is shown for two choices of the explosion energy $E_{\rm SN}$ and $^{56}$Ni mass $M_{\rm Ni}$ as listed in the legend, and the observed data is shown as black scatter points.}
    \label{fig:iPTF14gqr}
\end{figure}

Another potentially relevant event is the ultra-stripped SN iPTF14gqr \citep{De2018}, which was proposed to originate from a stripped star in a compact binary system. \cite{Wu2022} predicted that similar binary models for late-stage mass transfer as used in this work, in the mass range of $2.6$--$2.9\, M_{\odot}$ with $P_{\rm orb} \gtrsim 10$ d, could explain the inferred $M_{\rm csm} \approx 0.01\, M_{\odot}$ for iPTF14gqr. Figure \ref{fig:iPTF14gqr} shows that the initial shock cooling peak of our M2.75P30 model, which has $M_{\rm csm} \approx 0.02\, M_{\odot}$, resembles the early bolometric light curve of iPTF14gqr. 
With slightly larger values of the explosion energy and $^{56}$Ni mass, the second bump in the observed light curve could potentially be imitated by He SCE as well. However, iPTF14gqr was classified as Type Ic due to the absence of H or He absorption lines during the second peak, which may preclude the presence of an extended He envelope. 

Though the light curve morphology of our models is quite similar to the observed bolometric luminosity evolution of these fast-evolving events, comparison to other blackbody properties warrants further caveats. Firstly, we caution that the inferred blackbody radius and temperature for SN 2019kbj are discrepant with our model, despite the excellent agreement with the bolometric luminosity. \cite{Ben-Ami2023} infer the photosphere to be receding from a maximum radius of $\gtrsim 10^{15}$ cm throughout the duration of the plateau in the bolometric luminosity, whereas our model is still expanding over the first $\sim20$ d after explosion. 

In the case of iPTF14gqr, blackbody fitting of multiband photometry finds a temperature that matches our M2.75P30 model well during the first peak from CSM SCE. During the second light curve peak, the temperature appears to remain fairly constant. While our model also exhibits nearly constant color temperature, it is too hot by a factor of $\sim 1.25$ compared to the results from blackbody fitting of iPTF14gqr. We encounter a similar story for SN 2019dge, an ultra-stripped SN with two bumps in the bolometric light curve that \cite{haynie2025} found could be explained by a model with SCE from both the CSM and an extended He envelope \citep{Yao2020}. While the model replicates the high temperatures during the first peak from CSM SCE, the color temperature during the second bump is higher in our model than that inferred for SN 2019dge. 

Uncertainties in the estimated temperature exist for both the models and the observations. On the one hand, the color temperature in our models is typically set near $\sim 10^4$ K during the He envelope SCE phase due to He recombination, as we discuss above in Section \ref{sec:fiducialLCs}, but the models could be cooler if in reality the absorption opacity is larger than we have assumed. We discuss potential limitations of our current approach that may be causing our models to underestimate the absorption opacity in Section \ref{sec:uncertainty}. On the other hand, optical photometry bands tend to lie on the Rayleigh Jeans tail of the blackbody spectrum at the high temperatures $\gtrsim 10^4$ K found for these events and our models. If observed events have limited coverage in blue optical and UV bands, there may be some uncertainty in where the peak of the assumed blackbody lies, and fitting to multiband photometry may underestimate the true temperature. Both of these effects could bring theoretical and observed temperature predictions closer in line with each other. 

Of course, these events may not exhibit SCE from an extended He-rich envelope at all, with the second peaks in SN 2019dge and iPTF14gqr instead powered by $^{56}$Ni as originally assumed in the discovery papers \citep{De2018,Yao2020}. This is particularly likely for iPTF14gqr due to its classification as a Type Ic SN.
In our models, the temperature decreases below $10^4$ K once the light curve drops to the radioactive decay tail and the opacity is no longer set by He recombination, so observed epochs with lower temperatures may indicate that SCE is no longer dominant. Moreover, a binary system with an even tighter orbital period may be able to curtail the extent of the He envelope enough to suppress the signature of He envelope SCE in these ultra-stripped SNe \citep{Wu2022}.


\begin{figure*}
    \centering
    \includegraphics[width=\columnwidth]{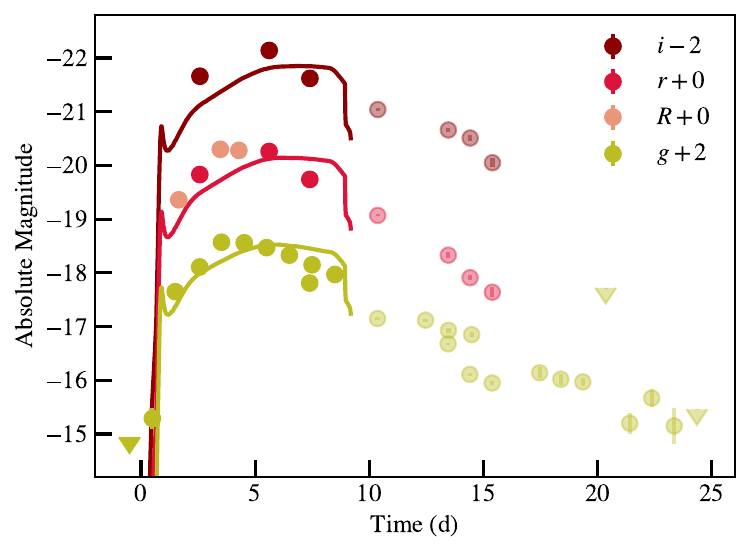}
    \includegraphics[width=\columnwidth]{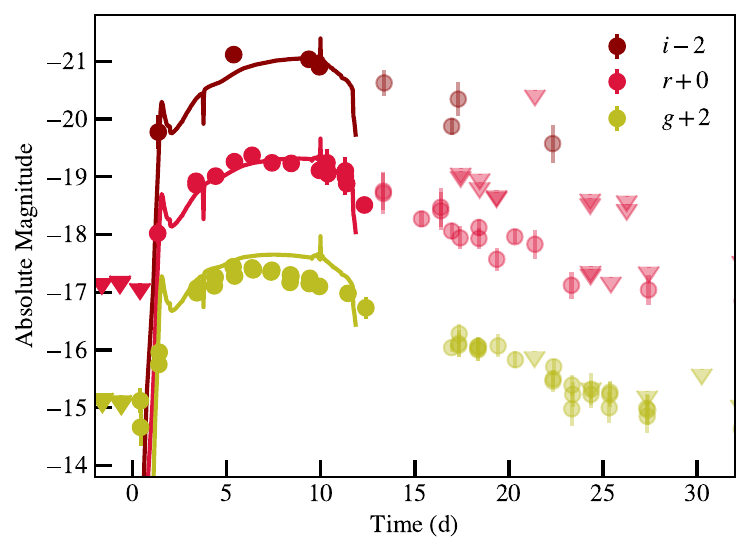}
    \includegraphics[width=\columnwidth]{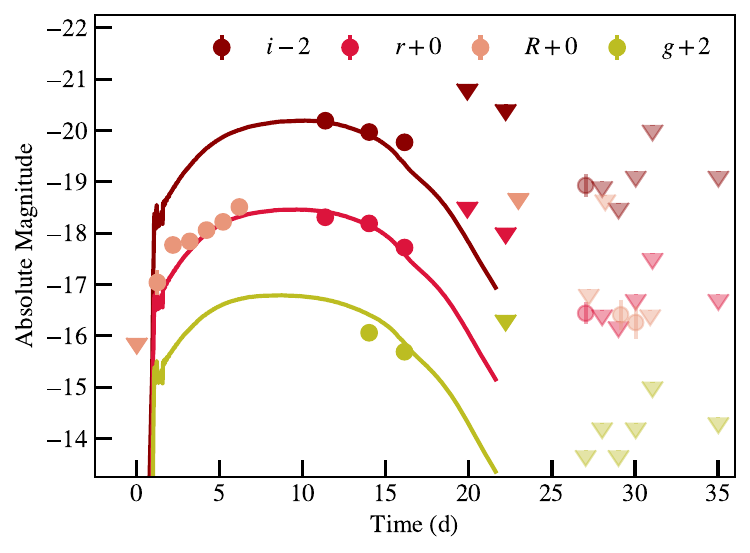}
    \includegraphics[width=\columnwidth]{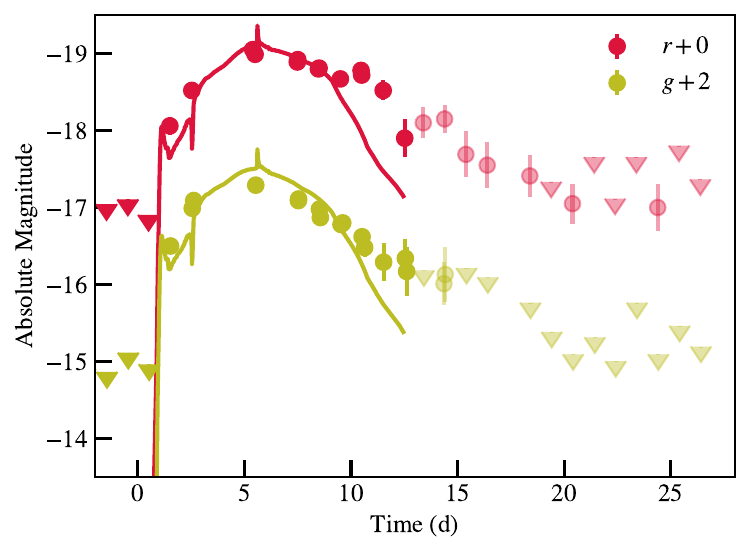}
    \includegraphics[width=\columnwidth]{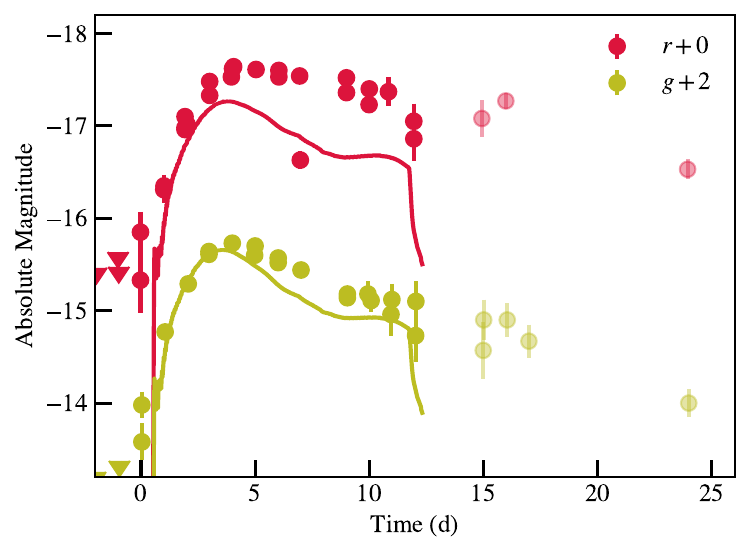}
    \includegraphics[width=\columnwidth]{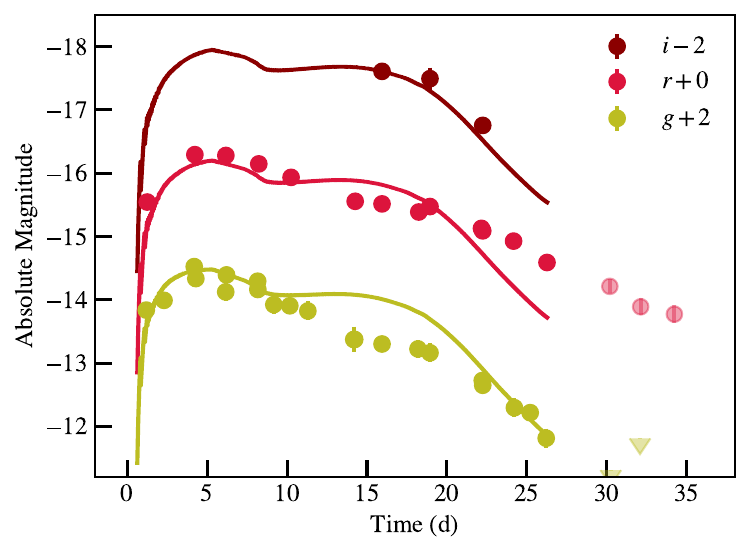}
    \caption{Comparison of the absolute magnitude in different bands for a selection of models and observed H-poor SNe. Observed data are scatter points, with upper limits indicated by upside-down triangles and colors denoting each band as listed in the legend. The models are shown as lines with the same colors for each observed band as the scatter points. \textit{Top left}: iPTF15ul \citep{Hosseinzadeh2017} with model M2.51P300 at $E=3 \times 10^{51}$ erg, $M_{\rm Ni}=0.05\, M_{\odot}$. \textit{Top right}: SN 2019myn \citep{ho2023} with M2.51P300 at $E_{\rm SN}=10^{51}$ erg, $M_{\rm Ni}=0.05\, M_{\odot}$. \textit{Center left}: iPTF15akq \citep{Hosseinzadeh2017} with model M2.51P500 at $E_{\rm SN}=10^{51}$ erg, $M_{\rm Ni}=0.05\, M_{\odot}$. \textit{Center right}: SN 2019php \citep{ho2023} with model M2.55P300 at $E_{\rm SN}=10^{51}$ erg, $M_{\rm Ni}=0.05\, M_{\odot}$. \textit{Bottom left}: SN 2018ghd \citep{ho2023} with model M2.65P100 at $E_{\rm SN}=4\times10^{50}$ erg, $M_{\rm Ni}=0.1\, M_{\odot}$. \textit{Bottom right}: SN 2023utc \citep{wang2025} with model M2.62P100 at $E_{\rm SN}=1.5\times10^{50}$ erg, $M_{\rm Ni}=0.1\, M_{\odot}$. } 
    \label{fig:observationalcomparisons}
\end{figure*}

\subsection{Absolute magnitude in optical bands}
\label{sec:absmagcomparisons}
We can also predict how the light curves of our models will appear in different observed bands. Figure \ref{fig:observationalcomparisons} shows a selection of comparisons between the light curves of our models and observed Type Ib/n SNe in \textit{g}-band and \textit{r}-band, along with \textit{i}-band and \textit{R}-band where available. The different bands are offset by arbitrary constants as listed in the legend, with observed data presented as scatter points and models shown as lines. 

SNEC assumes a blackbody spectrum at the color temperature and uses bolometric corrections to estimate the absolute magnitude in each filter. Since the blackbody assumption breaks down once the material in the SNEC simulation is no longer thermalized anywhere, the model light curves in each band end before the bolometric light curves. This tends to occur as the model transitions to the radioactive decay tail, so the comparisons focus on SCE in these events. Note that the sudden downturn at the end of some of the light curves may be exaggerated, as the temperature during the transition to the radioactive decay tail is sensitive to our choices for the amount and mixing of $^{56}$Ni. For times where our model absolute magnitudes are unknown, we show the model data as lighter shaded points.

The light curves in Figure \ref{fig:observationalcomparisons} exhibit a range of morphologies. Some rise rapidly to very bright peaks of $-18$--$-20$ mag in \textit{r}-band that last for $\sim 10$--$20$ d before declining. This behavior originates from models with large $\Porb\gtrsim 300$ d and low $\Mhe \lesssim 2.6\, M_{\odot}$, which expand significantly during their evolution and are highly stripped at late times. The ensuing SCE plateau is generated by contributions from both the very extended, massive CSM of $M_{\rm csm}\approx 0.3$--$0.5\, M_{\odot}$ and the smaller, but nearly equal mass, He envelope. 

In the top left panel of Figure \ref{fig:observationalcomparisons}, we see that the first $\sim10$ d of the H-poor SN iPTF15ul \citep{Hosseinzadeh2017} exhibits a light curve morphology comparable to the M2.51P300 model with $E_{\rm SN}=3\times10^{51}$ erg, $M_{\rm Ni}=0.05\, M_{\odot}$. For the same M2.51P300 model with $E_{\rm SN}=10^{51}$ erg instead, the top right panel shows how the model light curve resembles the first $\sim 10$ d of SN 2019myn \citep{ho2023}. These are both bright, fast-evolving events, best fit with CSM SCE that is enhanced by relatively high explosion energies.

The M2.51P500 model with $E_{\rm SN}=10^{51}$ erg, $M_{\rm Ni}=0.05\, M_{\odot}$ exhibits a broad, bright peak in optical bands, associated with a long plateau in the bolometric luminosity over $\sim 20$ d from combined CSM and He envelope SCE. This slow but bright light curve evolution is seen in the first $\sim 20$ d of iPTF15akq \citep{Hosseinzadeh2017}. Another luminous model, M2.55P300 with $E_{\rm SN}=10^{51}$ erg, $M_{\rm Ni}=0.05\, M_{\odot}$, reaches a peak as bright as SN 2019php \citep{ho2021} and appears to evolve similarly for the first $\sim 10$ d.



Some events with lower peak magnitudes look more like models with lower explosion energies, as well as CSM masses closer to $\sim 0.1\, M_{\odot}$. Compared to SN 2018ghd \citep{ho2023}, the \textit{g}-band light curve of the  M2.65P100 model with $E_{\rm SN}=4\times10^{50}$ erg, $M_{\rm Ni}=0.1\, M_{\odot}$ has a similar shape and duration out to $\sim 15$ d.  We note that the $r$-band light curve of SN 2018ghd is much flatter than both the observed $g$-band and our model light curves, which may indicate the influence of other effects such as extinction. The two bumps in the light curve of the M2.62P100 model using $E_{\rm SN}=1.5\times10^{50}$ erg, $M_{\rm Ni}=0.1\, M_{\odot}$ are reminiscent of SN 2023utc \citep{wang2025}. 


Considering the models discussed thus far, we see that nuances in binary stellar evolution can lead to light curves that span quite a few different morphologies, which are reflected in the diversity among observed events. 
Nevertheless, the above comparisons are heavily biased towards both observed and model light curves that remain bright on relatively long timescales. The models with these characteristics typically have $\Porb \geq 100$ d and $M_{\rm csm} \gtrsim 0.1\, M_{\odot}$. 

\begin{figure*}
    \centering
    \includegraphics[width=0.495\textwidth]{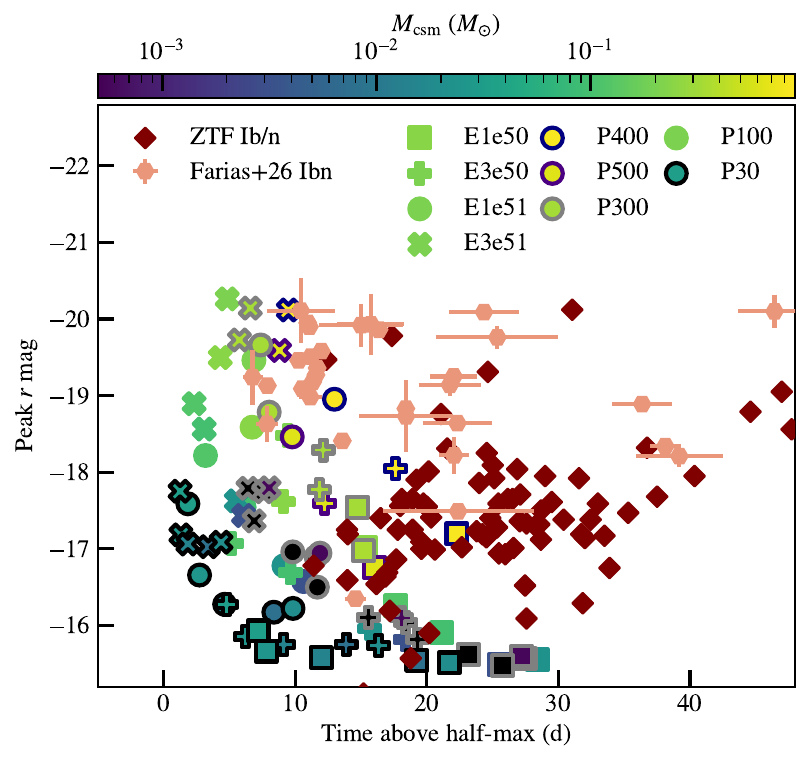}
    \includegraphics[width=0.495\textwidth]{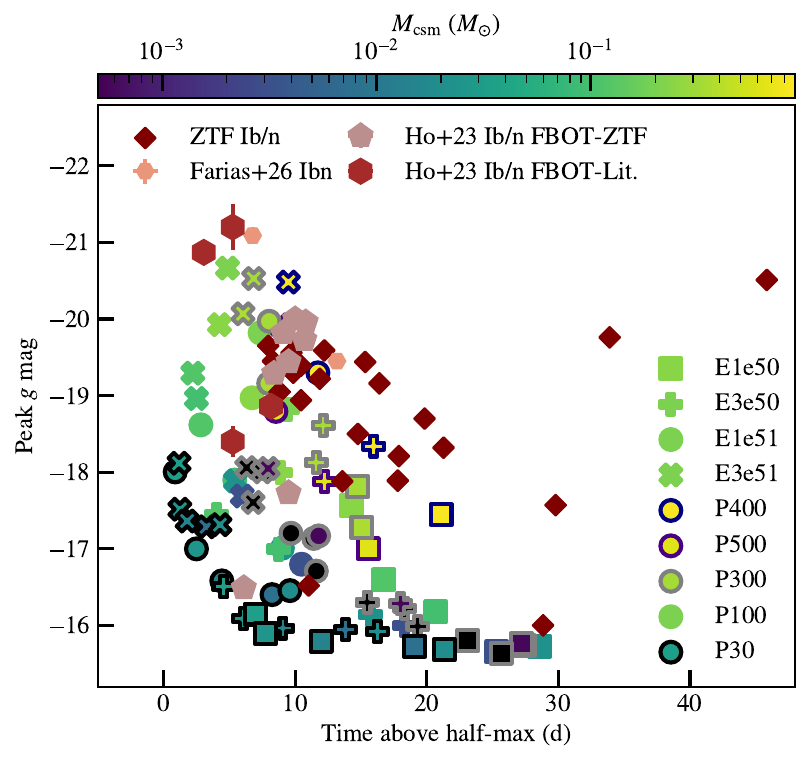}
    \caption{\textit{Left}: The peak $r$-band absolute magnitude versus the time above half-max in the $r$-band, shown for all models with varying explosion energy. Different explosion energies are denoted by distinct scatter point shapes: squares are models with $E_{\rm SN}=10^{50}$ erg, plus signs are $E_{\rm SN}=3\times 10^{50}$ erg, circles are $E_{\rm SN}=10^{51}$ erg, and crosses are $E_{\rm SN}=3\times10^{51}$ erg. The orbital period of each model is also denoted by the color of the scatter point outline, as listed in the legend, and the fill color of each scatter point corresponds to the value of $M_{\rm csm}$ in the color bar. Observed Type Ibn SNe from \cite{farias2026} and Type Ib/n SNe detected in the ZTF BTS \citep{perley2020} are also plotted over the models. \textit{Right}: Same as on the left, but for the $g$-band instead. Here we include observed Type Ib/n SNe classified as FBOTs from \cite{ho2023}, Type Ibn SNe from \cite{farias2026}, and Type Ib/n SNe detected in the ZTF BTS \citep{perley2020}.  }
    \label{fig:allmodelprops_obscomparison}
\end{figure*}

The other model light curves that are not represented in Figure \ref{fig:observationalcomparisons} come from systems with smaller CSM masses. These light curves feature a rapidly-evolving initial peak from CSM SCE, followed by a much dimmer phase of He envelope SCE. 
Similar events in the literature for Type Ib/n appear to be less common (see Figure \ref{fig:allmodelprops_obscomparison}), which may be expected given that the properties of these light curves are more difficult to capture. Higher cadence, deeper surveys, such as the Vera C. Rubin Observatory’s Legacy Survey of Space and Time \citep[LSST;][]{Ivezic19}, could probe whether these explosion models are represented in nature.

We note that some of the events we compare to in this section were also studied in \cite{farias2026},
which selected well-sampled Type Ibn SNe to model with MOSFiT \citep{guillochon2018} using CSM-SN ejecta interaction models and $^{56}$Ni radioactive decay. For SN 2019kbj, they infer $M_{\rm csm} \approx 0.04^{+0.014}_{-0.01}\, M_{\odot}$ and $R_{\rm csm} \approx 0.62^{+0.34}_{-0.18} \times 10^{14}$ cm,
and for iPTF15ul, they infer $M_{\rm csm} \approx 0.195^{+0.226}_{-0.094}\, M_{\odot}$ and $R_{\rm csm} \approx 6.62^{+6.64}_{-3.19}\times 10^{14}$ cm. 
The MOSFiT modeling favors less massive and confined CSM for SN 2019kbj than the M2.51P400 model shown in Figure \ref{fig:sn2019kbj}. For iPTF15ul, the CSM properties of the M2.51P300 model we use are within an order-of-magnitude of the MOSFiT values, though other inferred parameters like the ejecta mass are dissimilar. Our detailed numerical models can capture similar light curve properties to the observed events, but with different CSM and ejecta distributions, indicating that alternate scenarios to pure CSM-SN ejecta interaction may be worth considering when modeling these systems.

As discussed in Section \ref{sec:bolometriccomparisons}, uncertainties in the color temperature of our models may affect our prediction for the flux in each observed band. Since the \textit{g}- and \textit{r}-bands lie on the Rayleigh Jeans tail of the assumed blackbody, a cooler temperature would increase the flux in these bands.  Furthermore, our approach to generating the absolute magnitude in each filter is too simplistic to capture the effects of lines or extinction, which would affect how our models appear compared to real, observed SNe. 

Despite these uncertainties, we see from the comparisons in Figure \ref{fig:observationalcomparisons} that the shapes and magnitudes of the model light curves are similar to observed H-poor SNe. More careful investigation would be needed to understand whether the models shown ought to explain these specific events. Nevertheless, the resemblance shared between the models and the observations does illustrate that explosions of stripped stars in binaries like the progenitor systems modeled in this work can be observed in nature.

\subsection{Light curve properties}

Figure \ref{fig:allmodelprops_obscomparison} shows how the peak absolute magnitude in our model light curves trends with the time above half-max, defined as the amount of time that the light curve spends above half of the maximum absolute magnitude in a given band. We plot our models with fixed $M_{\rm Ni}=0.05\, M_{\odot}$ while varying the explosion energy from $E_{\rm SN} = 10^{50}$--$3\times10^{51}$ erg, as indicated by the shape of the scatter point in the legend. Model scatter points are shaded according to the value of $M_{\rm csm}$ in the color bar, and the points are outlined with different colors according to each model's $\Porb$, as listed in the legend.

In the left panel of Figure \ref{fig:allmodelprops_obscomparison}, we plot peak magnitude versus time above half-max for the $r$-band. Focusing on the effect of varying CSM mass, we see that the scatter points are mostly distributed from models with larger $M_{\rm csm} \gtrsim 0.1\, M_{\odot}$ (yellow/green points) at brighter peak magnitudes and systematically longer timescales, to models with lower CSM masses (blue/purple points) at dimmer peak magnitudes and generally shorter timescales. This trend arises because SCE dominates the light curves -- models with more massive CSM have longer diffusion timescales during SCE and also tend to be brighter (Equations \ref{eq:td} and \ref{eq:lumSCE}).

To isolate the effect of explosion energy at fixed $M_{\rm csm}$, we can compare points with similar colors, but different shapes. The explosion energy plays an important role in determining the peak magnitude and time above half-max for the light curve. The longest timescale light curves come from the lowest explosion energies of $E_{\rm SN} = 10^{50}$ erg (squares), as the slower shock velocities allow the model to evolve more slowly; these models also tend to be the least bright. In contrast, the highest explosion energies (crosses) occupy the upper left side of the plot, with very short peak timescales $\lesssim 10$ d and with the brightest peak magnitudes for a given CSM mass.

In addition to the properties of the models, we plot the estimated peak $r$-band magnitude and time above half-max for the light curves of a sample of Type Ibn SNe from \cite{farias2026} (salmon hexagons). These observed SNe tend to be brighter than $-18$ mag at $r$-band peak and are distributed across a wide spread in light curve timescale, with some not even shown within the figure limits. However, not many are observed with peak timescales $\lesssim 10$ d.  A sample of Type Ib/n SNe from the ZTF Bright Transient Survey\footnote{https://sites.astro.caltech.edu/ztf/bts/explorer.php} (BTS; red diamonds, without error bars; \citealt{perley2020}) contains less luminous events overall, but still with a large spread in light curve durations.

Compared to \cite{farias2026}, our models share properties with the brightest and fastest-evolving of these observed SNe Ibn. The most similar models come from the most highly stripped stars, with smaller ejecta masses $M_{\rm ej}$ and comparable $M_{\rm csm} \gtrsim M_{\rm ej}$. Since our models with $M_{\rm csm} \lesssim 0.01\, M_{\odot}$ tend to be both dimmer and more slowly-evolving, the observed SNe Ibn are systematically more luminous than these models with $M_{\rm ej} \gg M_{\rm csm}$. Only one event in the \cite{farias2026} sample lies in a region populated by the less luminous, lower $M_{\rm csm}$ models. As all but one of the events in Sections \ref{sec:bolometriccomparisons} and \ref{sec:absmagcomparisons} are classified as Type Ibn, this confirms why the light curve comparisons in those sections are biased towards models with more massive CSM. 

The ZTF BTS sample, which inclues both Type Ib and Type Ibn SNe, spans down to peak $r$-mag $\sim -16$--$-17$ and overlaps with some of our models with lower $M_{\rm csm} \lesssim 0.01\, M_{\odot}$ and lower energy explosions $E_{\rm SN} < 3\times 10^{50}$ erg. In particular, some of the models with $\Porb = 300$ d that have no nearby CSM exhibit similar light curve properties to the least luminous of the Type Ib SNe from ZTF.

In the right panel of Figure \ref{fig:allmodelprops_obscomparison}, we plot the peak magnitude versus time above half-max for the $g$-band. The overall distribution of our models exhibit the same trends as in the left panel, though the $g$-band light curves tend to be a little brighter than the $r$-band. As \citep{farias2026} focus their analysis on the $r$-band, there are fewer Type Ibn SNe with $g$-band properties compiled in this source, but we include what is available from \cite{farias2026} and the ZTF BTS Type Ib/n SNe. 
In $g$-band, only two events from these samples lie among models with $M_{\rm csm} \lesssim 0.01\, M_{\odot}$; a handful have similar properties to larger $M_{\rm csm} \gtrsim 0.1\, M_{\odot}$ models; and many are too bright and slowly-evolving to overlap with any of our models.

We note that in Figure \ref{fig:observationalcomparisons}, a few of the events used as comparisons are considered to be fast blue optical transients (FBOTs). Motivated by this, we compare the model properties to FBOTS classified as Type Ib or Type Ibn SNe from \cite{ho2023} (red hexagons and light brown pentagons, respectively). Some of these events are also included in the \cite{farias2026} sample in the left panel, so the effect of focusing on these FBOTs is mainly to isolate Type Ib/n SNe that were characterized as fast-evolving and blue. 

Perhaps unsurprisingly, the properties of the Type Ib/n FBOTs are well-described by the diversity in our model light curves. Many of our models fall under the FBOT definition: the majority spend fewer than $15$ d above half-max, unless exploded with lower energies; furthermore, all models tend to remain quite hot, and therefore blue, throughout the SCE phases of the light curves. Overall, we infer from this comparison that the explosions of stripped stars in binary systems like those modeled in this work likely contribute to the population of observed FBOTs, especially those that are spectroscopically classified as Type Ib/n SNe.

\section{Discussion and Conclusions}
\label{sec:discussion}

\subsection{Spectral features}
\label{sec:spectra}

We expect that the models will appear spectroscopically similar to Type Ib and Type Ibn SNe, as the models are devoid of H. For $M_{\rm csm} \lesssim 5\times 10^{-3}\, M_{\odot}$, the models essentially describe the explosions of stripped He stars without significant interaction, so they are unlikely to exhibit narrow line features. In contrast, the light curves of models with $M_{\rm csm} \gtrsim 0.1\, M_{\odot}$ are typically dominated by SCE from CSM and are likely to show signatures of interaction with He-rich CSM. For instance, although we removed this material from the SNEC simulations for numerical reasons, cool, low-density CSM located at larger radii can become ionized by the shock breakout. Before it is swept up by the shock, this could produce narrow emission lines at very early times, even before the SCE peak in the light curve, as so-called flash-ionization features \citep[e.g.,][]{khazov2016}. 

Our least massive models can have very high luminosities during SCE with $L \gtrsim 10^{44}$ erg/s (Figure \ref{fig:fiducialLCs}), and they are also very hot during this phase ($T\gtrsim 2\times 10^4$ K). It is possible that these explosions can appear initially featureless before the luminosity declines enough for line formation \citep{aspegren2026}, which may affect spectroscopic classification at peak. Another consideration is the CSM velocity, which affects the width of observed spectral lines. While Type Ibn SNe show characteristic narrow emission lines from unshocked CSM that can persist over many days \citep{farias2026}, this is not likely to occur in our models, since the shock typically breaks out from the edge of the CSM and accelerates all the CSM during breakout \citep[e.g.,][]{khatami2024}. Instead, the CSM will already be moving at the SN ejecta velocity and any He lines in the spectra will lack a narrow component. A spectroscopic classification of Type Ib instead of Type Ibn may therefore be more appropriate for many of our models.

Similar binary progenitors at shorter orbital periods $\Porb\lesssim 10$ d are likely to produce lower-density CSM profiles that allow for shock breakout at radii much smaller than the total CSM radius, as their mass loss rates are usually lower and mass can be ejected from earlier times \citep{Wu2022}. This would allow for the scenario of interior breakout, so that unshocked CSM ahead of the shock can be ionized and form the narrow lines that characterize the SNe Ibn class \citep[e.g.,][]{khatami2024}. Though the lower $\Porb$ models were not included in our model grid due to numerical difficulties, these systems can also produce dense nearby CSM and are very relevant to the population of H-poor interacting SNe.

\subsection{Associated radio emission and SN precursors}

During the lifetimes of the progenitors modeled in this work, the stripped stars experience multiple phases of mass loss, the most recent of which will produce the nearby CSM that affects the early optical light curve. The previous phase of mass transfer that occurs during C burning typically features mass loss rates of $\sim 10^{-5}$--$10^{-4}\, M_{\odot}\,\mathrm{yr}^{-1}$ over $10^3$--$10^4$ yr that can strip $\sim 0.1$--$1\, M_{\odot}$ from the star. Assuming non-conservative mass transfer where this material is lost from the binary system, the CSM from this earlier mass loss phase can be approximated as a dense wind out to $\sim 10^{18}$ cm. 

Given this distribution of dense CSM, interaction between the SN shock and the CSM can lead to luminous radio synchrotron emission rising  $\sim$ months--years after the SN explosion \citep{Stroh21,Rose24,wu2025}. As a result, it may be expected that the optical light curves shown in this work should accompanied by bright late-time radio emission. While prompt radio follow-up has been implemented for a subset of the FBOTs shown in Figure \ref{fig:observationalcomparisons}, these all resulted in non-detections \citep{ho2023}. The limits indicate radio emission no greater than $L_{\nu} \lesssim 10^{27}$--$10^{30}\, \mathrm{erg}\,\mathrm{s}^{-1}\, \mathrm{Hz}^{-1}$ during the first $\sim10$--$100$ d after the SN. 

Nevertheless, the non-detections as seen in \cite{ho2023} do not rule out the late-time radio emission from interaction between the SN shock and dense CSM from binary interaction that is predicted in \cite{wu2025}. Due to free-free absorption and synchrotron self-absorption, the 3 GHz radio light curves for the lowest-mass models in \cite{wu2025}, which are most relevant to this work, surpass $L_{\nu} \sim 10^{27}\, \mathrm{erg}\,\mathrm{s}^{-1}\, \mathrm{Hz}^{-1}$ only at $\sim 1$ yr after core collapse for the case of an asymmetric torus of CSM. Furthermore, \cite{wu2025} consider that the CSM may need to be accelerated in order to explain late-time VLASS detections \citep{Stroh21}. Faster CSM creates a cavity of low-density material for models with $\Mhe \lesssim 3\, M_{\odot}$, which is too diffuse for bright radio emission until a sharp rise at $\sim $1 yr for all wavelengths. This late-time radio behavior is not likely to be constrained from prompt radio follow-up in the first year after the SN.

In most of our models, the late-time mass transfer phase that produces nearby CSM and bright shock cooling in the early optical light curve is preceded by the C burning mass transfer phase that can produce dense CSM at $\sim 10^{17}$--$10^{18}$ cm. Only the largest $\Porb\gtrsim 400$ d models and the $\Mhe\sim 2.7\, M_{\odot}$, $\Porb=300$ d models avoid Roche lobe overflow during C burning. As a result, the binary interaction scenario for producing nearby CSM explored in this work is strongly associated with late-time radio emission powered via SN shock interaction with the more extended CSM. Future late-time radio follow up of Type Ib/n SNe with signatures of SCE in the optical light curve will provide strong constraints on the viability of this binary scenario to explain the multiwavelength signatures of interacting SNe. If radio emission is not detected, it may indicate that mass transfer during C burning is at least partially conservative, such that not all of the stripped material is ejected from the system to form a circumbinary torus.


The extreme mass transfer before the SN predicted by the binary evolution models may also give rise to precursor emission for Type Ib/n SNe \citep[e.g.,][]{Pastorello2007,strotjohann2021}.
For instance, accretion of the mass stripped from the donor star onto a compact object companion could produce an accretion disk around the companion, as well as a circumbinary outflow (CBO) composed of material lost from the binary system. Interaction between the disk wind and the CBO may produce optical emission with luminosities comparable to observed precursors in Type Ibn SNe \citep{Tsuna2024a}. If accretion onto the compact object produces an energetic wind or jet that emerges from the CBO, associated pre-SN X-ray or radio emission may also appear.

While we assume that mass transfer remains stable throughout our binary evolution models, the high rates of mass loss during the last few years of the binary system's lifetime may trigger unstable mass transfer \citep{Wu2022}. In that case, the low-mass He star could undergo runaway Roche lobe overflow, leading to a merger that could either trigger an explosion or become curtailed by core collapse of the SN progenitor. Given a compact object companion, super-Eddington accretion of the transferred mass may lead to an accretion disk wind \citep{Tsuna2024}. Emission from the disk wind, reprocessed into the optical/UV by the CBO, can power a rising precursor over $\sim$year timescales, which strongly resembles the precursor activity in SN 2023fyq \citep{dong2024}. Notably, the radio emission associated with SN 2023fyq can be explained by similar CSM density profiles to that of the CBO generated during the merger of the binary system \citep{Tsuna2024,baerway2025}.

\subsection{Current and future surveys}

Surveys with cadences of a few days or less, such as the Zwicky Transient Facility \citep[ZTF;][]{ZTF}, ULTRASAT \citep{Shvartzvald24}, and the Vera C. Rubin Observatory Legacy Survey of Space and Time \citep[LSST;][]{Ivezic19}, are well-suited to detect the explosions modeled in this work. For models with lower CSM masses, capturing the rapidly decline of the light curve from SCE requires a more frequent cadence. Models with massive CSM produce extended shock breakout over $\sim$day timescales that would also be interesting to study with sub-day cadence observations. 

In addition, comparison of our models to Type Ib/n SNe \citep{perley2020,farias2026} reveals that a large fraction of our models are dimmer and evolve more rapidly than typical Type Ib/n SNe. These progenitor models, which tend to have lower CSM masses, may exist in an unexplored region of parameter space that could be unveiled with the advent of deeper, high-cadence surveys such as LSST. In particular, the light curves of our progenitors in wide orbits that lack nearby CSM can also exhibit SCE from the extended He envelope, which is dimmer than the SCE from CSM by an order of magnitude. Yet detection of SCE in such events, likely as Type Ib SNe, would allow direct measurement of the radius of the stripped progenitor's envelope, thereby probing late-stage stellar evolution. 

As our models are very hot throughout the shock breakout and CSM SCE, these early light curve phases may be particularly suited for detection with upcoming UV telescopes like ULTRASAT. Photometric data in the UV will also provide better constraints on the temperature of the SN and will be particularly useful for these events that may hover at around $10^4$ K due to He recombination.
In the future, collecting UV spectra in the first hours of the SN through missions like UVEX will also reveal more details about the nearby CSM around these SN progenitors.

\subsection{Model uncertainties and future work}
\label{sec:uncertainty}

When constructing the CSM in our models, we assume an ejection velocity for the CSM which strongly determines the density and radial extent of the resultant CSM profile. Motivated by SPH simulations of mass loss from L2 \citep{pejcha2016}, we set $v_{\rm csm}\approx 0.4\, v_{\rm orb, c} \approx 20\, \mathrm{km}\ \mathrm{s}^{-1}\, (P_{\rm orb}/100\, \mathrm{d})^{-1/3}\, ((M_*+M_c)/5\, M_{\odot})^{1/3} $ (Section \ref{sec:massloss}). However, narrow lines associated with CSM that appear in the spectra of interacting SNe have typical line widths of $\sim 10^2$--$10^3~\mathrm{km}\ \mathrm{s}^{-1}$ \citep[][]{Pastorello08,Hosseinzadeh2017,strotjohann2021}, an order of magnitude larger than $v_{\rm csm}$ in our models. The CSM may be accelerated after leaving the system, for instance by radiation from the SN \citep{Tsuna2023} or even an accretion disk wind to velocities $\sim 10^{3}~\mathrm{km}\ \mathrm{s}^{-1}$ \citep{Tsuna2024}. 

In the latter case, the CSM could already be more diffuse and extended than we assume at the time of the explosion, which would affect our predictions for the explosion light curves. The shock could break out at smaller radii for less dense CSM (equivalently, a faster CSM velocity $v_w$ in Equation \ref{eq:r_d}) and lead to a phase of continued interaction before SCE, which would produce a different light curve shape \citep{Khatami2019}. For a larger CSM radius, the shock-cooling material could potentially be more extended and have a lower color temperature, affecting the predicted flux in optical bands. As the actual velocity of the CSM before the SN is not well-constrained, it would be worthwhile to explore these uncertainties in future work. 

In our SNEC simulations, we must choose where to distribute the $^{56}$Ni in the ejecta, opting to use a moderate mixing scheme where $^{56}$Ni is spread throughout the inner 50\% of the mass of the stellar ejecta. However, the literature on core-collapse SN models has not converged on the optimal decision for the strength of $^{56}$Ni mixing, with some allowing $^{56}$Ni to mix out to the edge of the stellar ejecta \citep{dessart2020}. The light curve shapes are sensitive to this choice \citep{Haynie2023}. We also smooth the composition profile in SNEC at the start of the simulation using a boxcar method with a certain width. In our models, we find that alternate choices for the $^{56}$Ni mixing boundary and the boxcar width cause minor differences to the light curve during SCE. 

Given the limitations of our methods, the color temperatures of our models are likely upper limits to the true photon temperature. Our calculation of color temperature implicitly assumes that the SED can be approximated by a single-temperature blackbody and is also highly sensitive to uncertainties in the opacity tables used in SNEC. When the properties of the ejecta fall outside of the limits of the opacity tables, whether in terms of composition, temperature, or density, our estimate of the absorption opacity is likely incorrect. 

Moreover, SNEC uses grey Rosseland mean opacities, which cannot capture the effects of frequency-dependent absorption opacity. In practice, photons of different frequencies may experience different amounts of absorption and remain in equilibrium with the gas temperature out to larger radii \citep{magee1995}. Furthermore, in optically thin regimes the Planck mean, which is more heavily weighted towards line absorption, may be more appropriate, and this can exceed the Rosseland mean by an order of magnitude \citep[e.g.,][]{takei2022}.
Future work ought to incorporate more realistic, frequency-dependent absorption opacities in order to more accurately predict the observed photon spectrum. 

\subsection{Conclusions}

The rapid and extreme expansion of low-mass stripped stars during their late nuclear burning phases leads naturally to the presence of dense CSM near the progenitors at core collapse, and this has been conjectured to explain the origins of H-poor interacting SNe.
In this work, we explore the appearance of SN explosions from stripped star progenitors with nearby CSM, produced by late-time mass transfer in binary systems. 

To generate the CSM density profiles, we use MESA binary stellar evolution simulations to track the mass loss rates from low-mass stripped stars with a range of initial He star masses, $\Mhe \sim 2.5$--$2.75\, M_{\odot}$, placed in binaries of different initial orbital periods, $\Porb \sim 30$--$500$ d. The resulting CSM is lost at rates of $10^{-3}$--$10^{-1}\, M_{\odot}\, \mathrm{yr}^{-1}$ in the last few months--years before core collapse, leading to total CSM masses $\sim 10^{-3}$--$10^{-1}\, M_{\odot}$ extending to radii of $\sim 10^{13}$--$10^{14}$ cm. 

We explode the progenitors in SNEC to study their SN light curves. The model light curves are dominated by phases of shock cooling emission from both the dense CSM and the He envelope of the stripped star, which tends to be fairly extended. For models where the CSM mass is comparable to the mass of the star's He envelope, a bright plateau from SCE is visible in the bolometric light curves that drops sharply to the radioactive decay tail. The light curves of models with lower CSM masses display rapidly evolving initial peaks from CSM SCE, followed by longer-timescale, dimmer SCE from the He envelope. Even models with negligible CSM exhibit He envelope SCE, so that they are brighter than a light curve powered only by radioactive decay. 

The appearances of our model light curves are reflected in the population of known Type Ib/n SNe, which include events that manifest extremely bright and fast-evolving peaks similar to SCE from $M_{\rm csm}\gtrsim 0.1\, M_{\odot}$. Broader, less luminous light curves can be achieved with lower explosion energies and larger ejecta masses. Some of the models emulate the fastest and brightest of Type Ib/n SNe, but our full model grid includes even dimmer and more rapidly-evolving light curves than most Type Ib/n SNe. Overall, the luminosities and temperatures of our models can be characterized as fast and blue, and the properties of the model grid overlap significantly with the population of FBOTs. 

The advent of more sensitive, high cadence surveys, such as LSST, will enable detection of less luminous transients. 
As more SNe are discovered in the future, it will be interesting to look for signatures of SCE from the extended He envelope, as this phase can probe the radius and mass of the SN progenitor itself. 
Identification of this He envelope SCE phase in observed SNe would provide a window into the late-time evolution of binary SN progenitors. 

In the future, we plan to explore the colors and spectra of our numerical light curve models using more detailed treatments of the radiative transfer, particularly related to the effects of frequency-dependent opacities. We hope thereby to progress towards a more robust understanding of the observed photon energy distribution in these SNe. In addition, our modeling framework can be used to explore other potential progenitor systems for interacting SNe, such as systems that host more massive progenitors or that experience unstable mass transfer.

\begin{acknowledgments}
    We thank Xiaoshan Huang, Wenbin Lu, Brenna Mockler, Daichi Tsuna, and Sunny Wong for helpful discussions.
\end{acknowledgments}

\appendix
\section{Color temperature in SNEC}
\label{sec:appendixcolorTeff}
Previously, SNEC measured the temperature at the location of the photosphere and assumed that this effective temperature represented the blackbody temperature for the photon SED \citep{SNEC}. This only affects the computation of the absolute magnitudes in different observed bands, which uses a bolometric correction table and assumes a blackbody approach without effects from lines. While still operating under the blackbody framework, we can improve our estimate for the temperature of the photons. 

\begin{figure}
    \centering
    \includegraphics[width=0.45\textwidth]{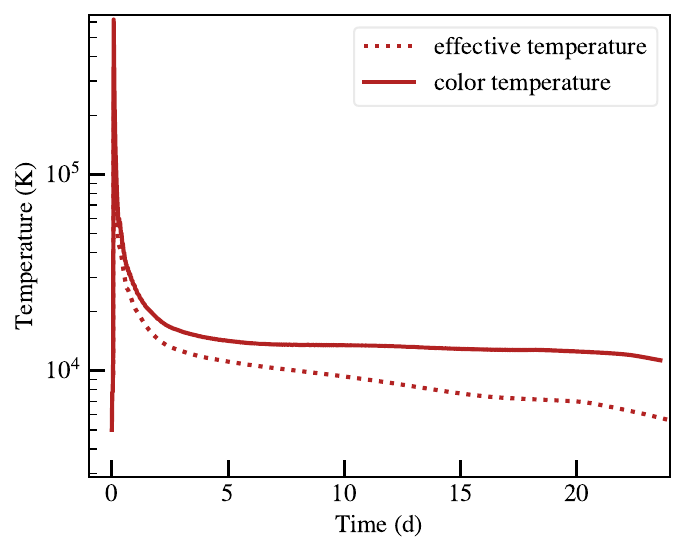}
    \includegraphics[width=0.45\textwidth]{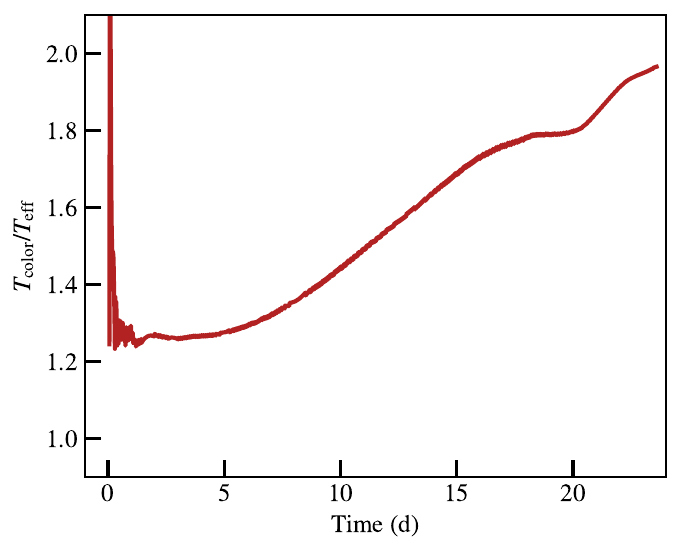}
    \caption{Comparison of the color temperature to the effective temperature in M2.75P100. Throughout the evolution out to $\sim 25$ d, the color temperature as calculated with the methods of Section \ref{sec:colortemperature} ranges between 1.2--2 times larger than the effective temperature. Much of the difference occurs because the color temperature remains fairly constant during the He envelope shock cooling emission phase, due to the increase in opacity at $\sim 10^4$ K from He recombination. }
    \label{fig:appendixTcolorTeff}
\end{figure}

As described in Section \ref{sec:colortemperature}, the color temperature of the photons is likely set at the location where the photons are last in thermal equilibrium with the gas, called the thermalization radius. Figure \ref{fig:appendixTcolorTeff} shows the difference between the two methods of calculating the temperature of the photons, the effective temperature $T_{\rm eff}$ and the color temperature $T_{\rm color}$, for a selected model (M2.75P100). The effective temperature is systematically lower than the color temperature and declines more rapidly. The color temperature remains quite constant because the location of the thermalization depth is set by a jump in opacity due to He recombination, which takes place at $\sim 10^4$ K -- this sets the color temperature until the entire He envelope recombines.

The ratio $T_{\rm color}/T_{\rm eff}$ is at a minimum $\sim 1.2$ and at most $\sim 2$. This is representative of most models, although the models with $M_{\rm csm} \sim M_{\rm ej}$ tend to have color temperatures that remain even hotter over the same timescale. After the color temperature curve ends, the model is no longer thermalized at any location and the blackbody approach certainly breaks down. Typically, the contribution of $^{56}$Ni begins to dominate once helium fully recombines and the entire ejecta is optically thin, so the color curves end about when the shock cooling emission becomes subdominant to radioactive decay.

\begin{figure}
    \centering
    \includegraphics[width=0.45\textwidth]{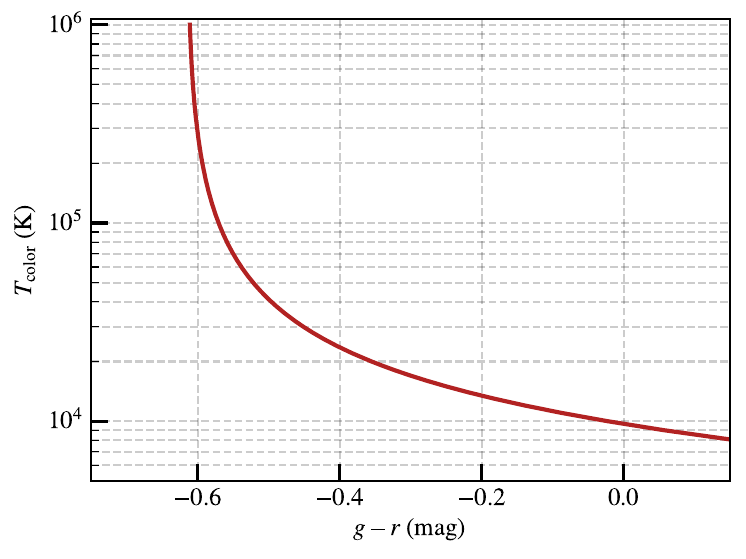}
    \caption{The relationship between the assumed blackbody color temperature and the inferred $g-r$ color for the bolometric correction methods in SNEC.}
    \label{fig:appendixT_vs_color}
\end{figure}
Figure \ref{fig:appendixT_vs_color} shows how the choice of the color temperature affects the predicted $g-r$ color according to the methods used in SNEC, namely the bolometric correction table. A change from using a blackbody color temperature of $\lesssim 10^4$~K to a color temperature that is a factor of 2 larger can lead to a difference in $g-r$ color of $\sim - 0.2$ mag. As a result, the correction to use the color temperature at the thermalization depth instead of the photosphere temperature can have significant effects on the predicted observational appearance of the explosion models in SNEC, especially for these stripped star progenitors that remain quite hot due to helium recombination.

\section{Varying explosion energy}
\label{appendix:energy}
\begin{figure*}
    \centering
    \includegraphics[width=0.32\textwidth]{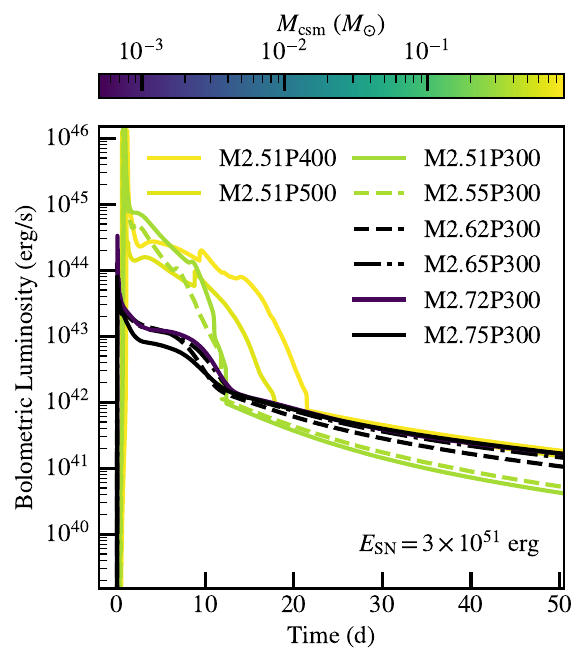}
    \includegraphics[width=0.32\textwidth]{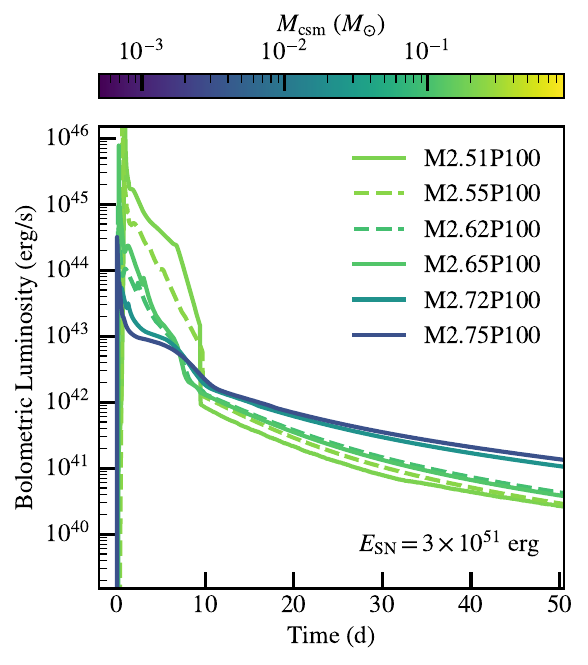}
    \includegraphics[width=0.32\textwidth]{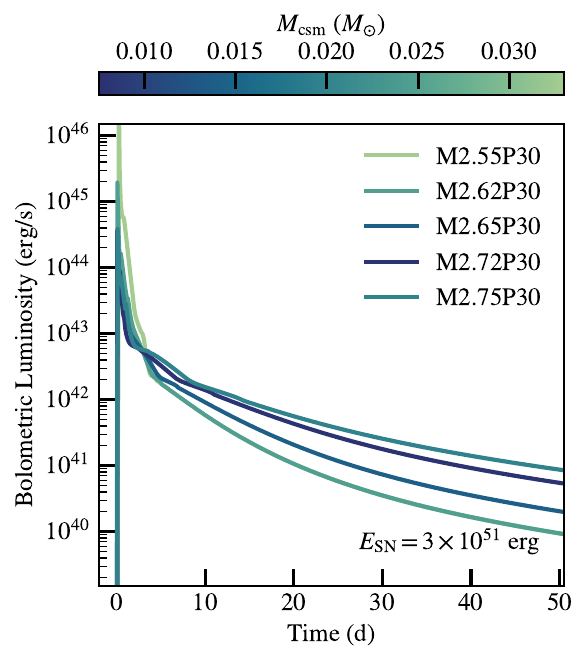}
    \includegraphics[width=0.32\textwidth]{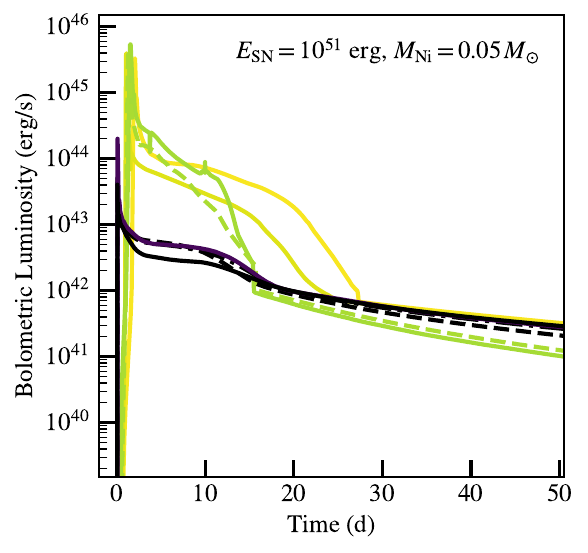}
    \includegraphics[width=0.32\textwidth]{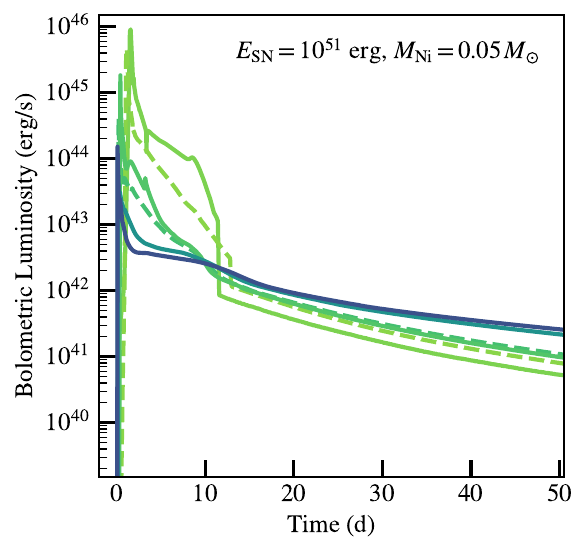}
    \includegraphics[width=0.32\textwidth]{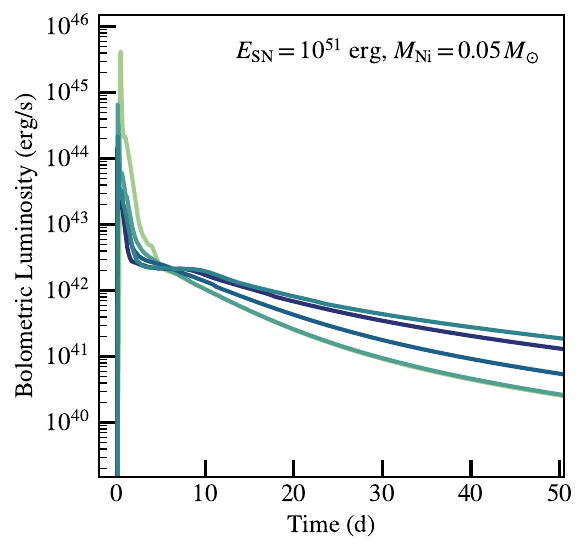}
    \includegraphics[width=0.32\textwidth]{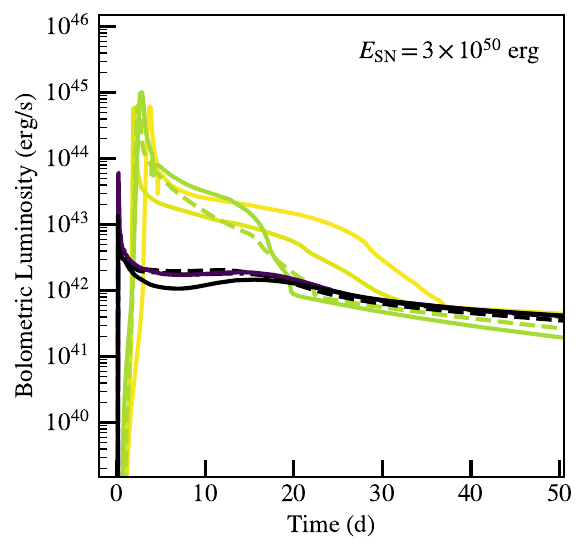}
    \includegraphics[width=0.32\textwidth]{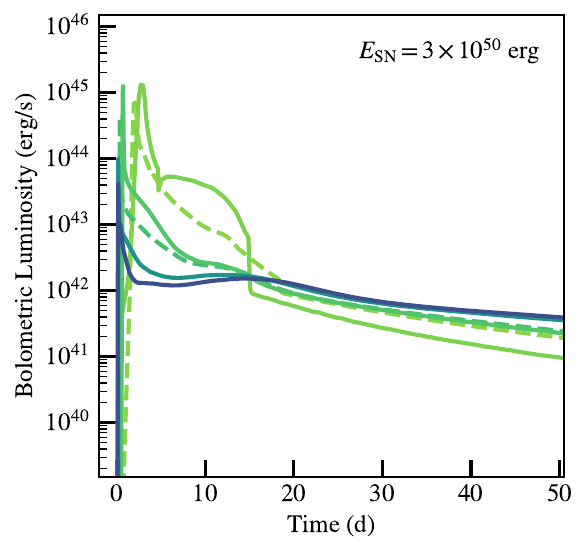}
    \includegraphics[width=0.32\textwidth]{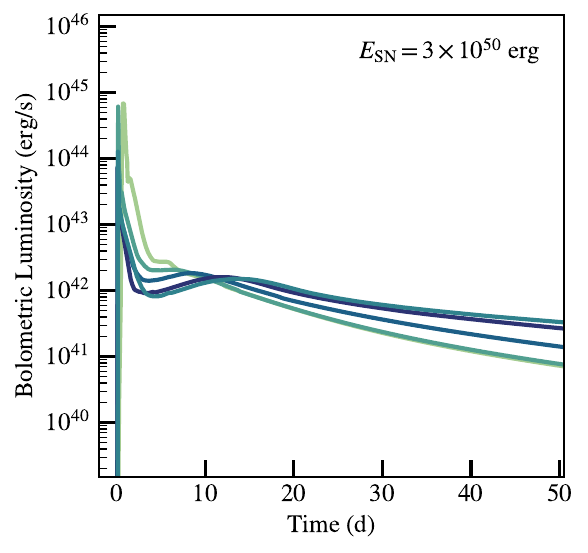}
    \includegraphics[width=0.32\textwidth]{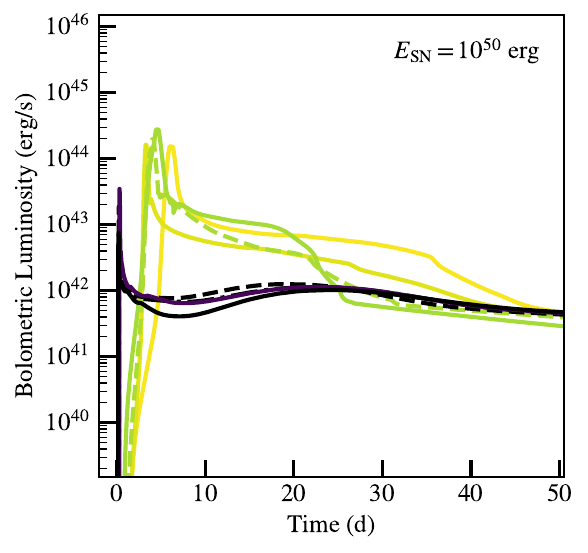}
    \includegraphics[width=0.32\textwidth]{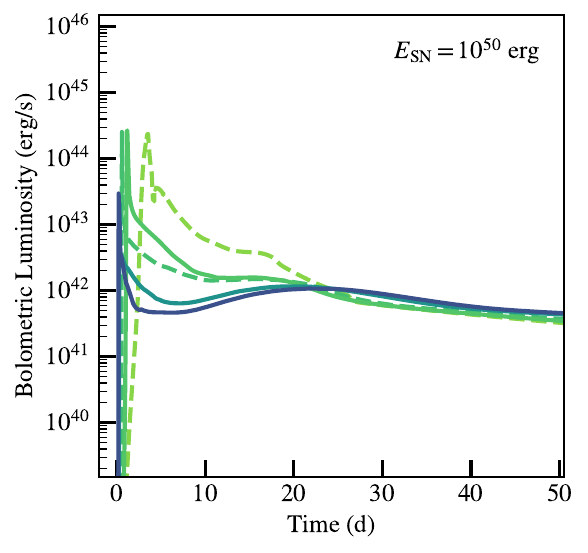}
    \includegraphics[width=0.32\textwidth]{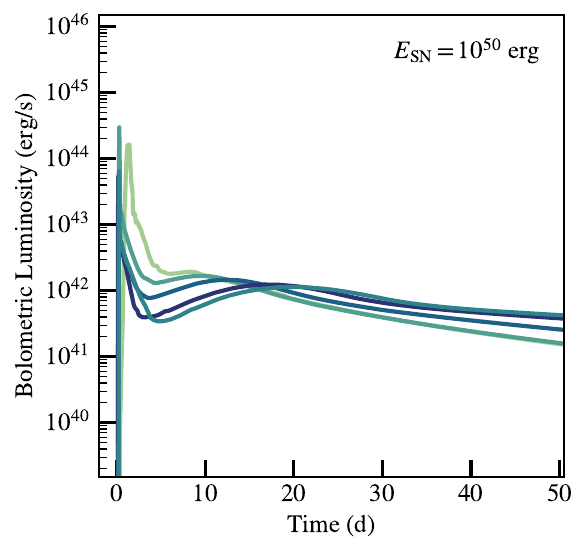}
 
    \caption{Model light curves for $M_{\rm Ni}=0.05\, M_{\odot}$, with varying $E_{\rm SN}$.}
    \label{fig:EvariedLCs}
\end{figure*}

Figure \ref{fig:EvariedLCs} shows all the models with different values of the explosion energy, $E_{\rm SN}$, and fixed $^{56}$Ni mass of $M_{\rm Ni}=0.05\, M_{\odot}$. From the top row to the bottom row, the explosion energy varies from larger to smaller values of $E_{\rm SN}=3\times10^{51}\, \mathrm{erg}$, $E_{\rm SN}= 10^{51}\, \mathrm{erg}$, $E_{\rm SN}= 3\times10^{50}\, \mathrm{erg}$, and $E_{\rm SN}= 10^{50}\, \mathrm{erg}$. Note that the M2.51P100 model is not shown for $E_{\rm SN}= 10^{50}\, \mathrm{erg}$ due to numerical issues.

\section{Varying nickel mass}
\label{appendix:Ni}
\begin{figure*}
    \centering
    \includegraphics[width=0.32\textwidth]{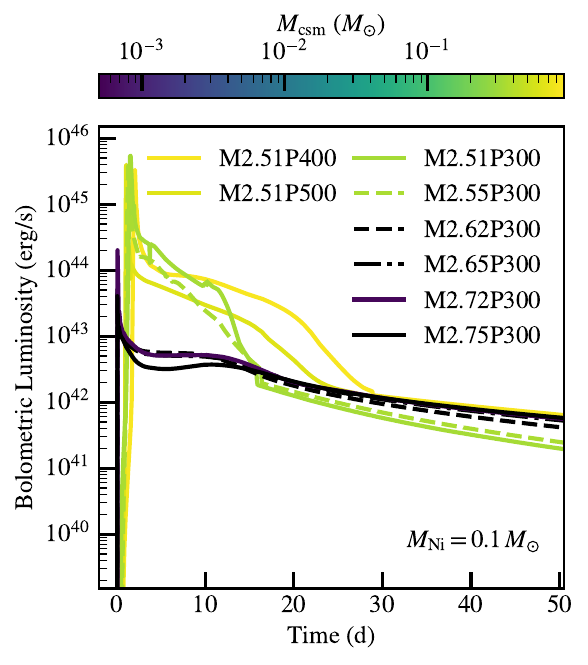}
    \includegraphics[width=0.32\textwidth]{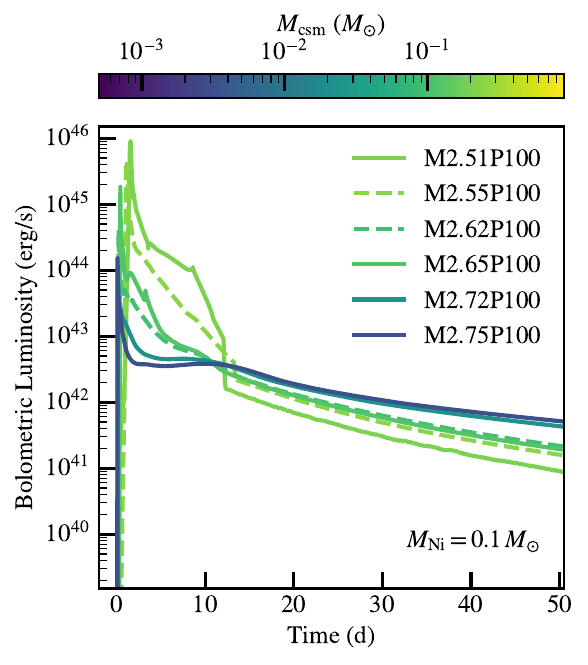}
    \includegraphics[width=0.32\textwidth]{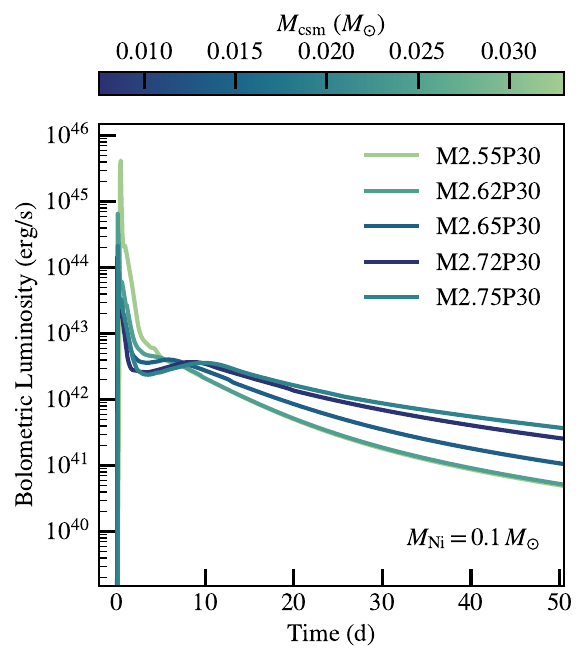}
    \includegraphics[width=0.32\textwidth]{large_Porb_fiducial_bolometricL_noCB.pdf}
    \includegraphics[width=0.32\textwidth]{P100d_fiducial_bolometricL_noCB.pdf}
    \includegraphics[width=0.32\textwidth]{P30d_fiducial_bolometricL_noCB.pdf}
    \includegraphics[width=0.32\textwidth]{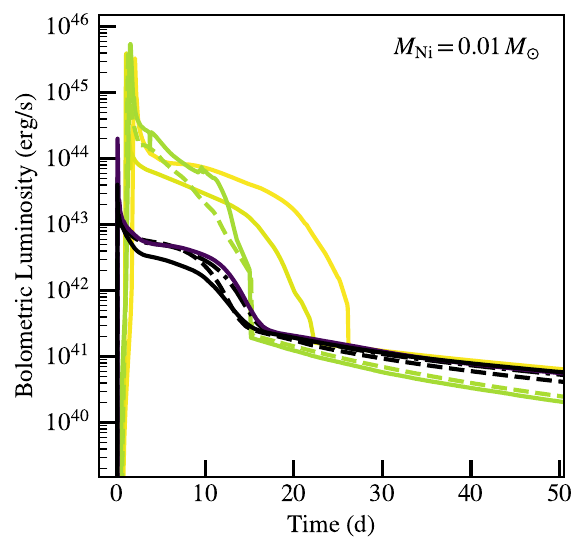}
    \includegraphics[width=0.32\textwidth]{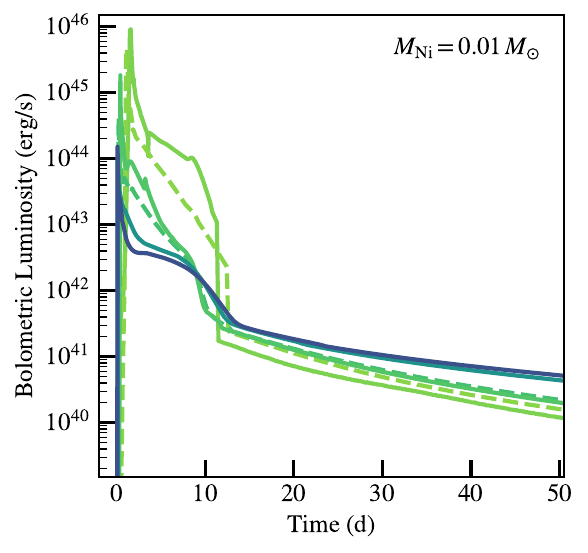}
    \includegraphics[width=0.32\textwidth]{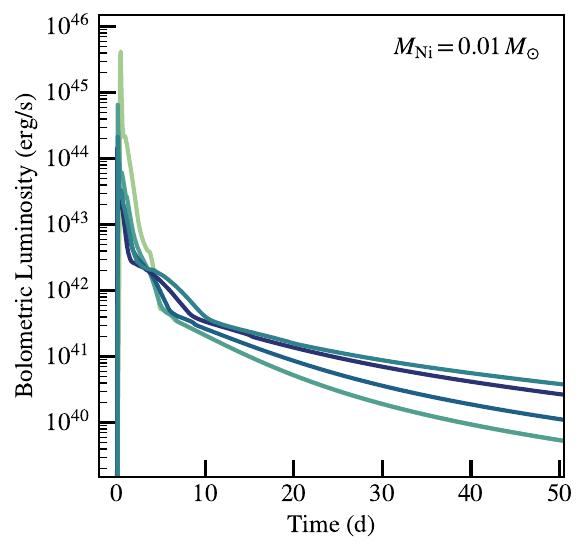}

    \caption{Model light curves for $E_{\rm SN} = 10^{51}\, \mathrm{erg}$, with varying $M_{\rm Ni}$.}
    \label{fig:NivariedLCs}
\end{figure*}

Figure \ref{fig:NivariedLCs} shows all the models with different values of the $^{56}$Ni mass and fixed explosion energy $E_{\rm SN}= 10^{51}\, \mathrm{erg}$. From the top row to the bottom row, the $^{56}$Ni mass varies from larger to smaller values of $M_{\rm Ni}=0.1\, M_{\odot}$, $M_{\rm Ni}=0.05\, M_{\odot}$, and $M_{\rm Ni}=0.01\, M_{\odot}$. The shape of the SCE phase changes with different $^{56}$Ni mass, becoming more prominent as $M_{\rm Ni}$ decreases, but the luminosity and timescale of the SCE is not greatly altered.

\bibliography{bib}
{}
\bibliographystyle{aasjournalv7}

\end{document}